\documentclass[twocolumn,twocolappendix]{aastex631}
\usepackage{amsmath} 
\usepackage{graphicx}
\usepackage{multirow}

\newcommand*{\hi}{\rm{H}\,\rm{\textsc{i}}}
\newcommand*{\msun}{\ensuremath{\rm{M}_{\odot}}}

\newcommand*{\kms}{\text{km}\,\text{s}\ensuremath{^{-1}}}

\defcitealias{bennet2018}{B18}
\defcitealias{jones2021}{J21}
\defcitealias{fielder2023}{F23}

\begin{document}

\title{All Puffed Up: Exploring Ultra-diffuse Galaxy Origins through Galaxy Interactions}

\correspondingauthor{Catherine Fielder}
\email{cfielder@arizona.edu}

\author[0000-0001-8245-779X]{Catherine Fielder}
\affiliation{Steward Observatory, University of Arizona, 933 North Cherry Avenue, Rm. N204, Tucson, AZ 85721-0065, USA}

\author[0000-0002-5434-4904]{Michael G. Jones}
\affiliation{Steward Observatory, University of Arizona, 933 North Cherry Avenue, Rm. N204, Tucson, AZ 85721-0065, USA}

\author[0000-0003-4102-380X]{David J. Sand}
\affiliation{Steward Observatory, University of Arizona, 933 North Cherry Avenue, Rm. N204, Tucson, AZ 85721-0065, USA}

\author[0000-0001-8354-7279]{Paul Bennet}
\affiliation{Space Telescope Science Institute, 3700 San Martin Drive, Baltimore, MD 21218, USA}

\author[0000-0002-1763-4128]{Denija Crnojevi\'{c}}
\affil{University of Tampa, 401 West Kennedy Boulevard, Tampa, FL 33606, USA}

\author[0000-0001-8855-3635]{Ananthan Karunakaran}
\affiliation{Department of Astronomy \& Astrophysics, University of Toronto, Toronto, ON M5S 3H4, Canada}
\affiliation{Dunlap Institute for Astronomy and Astrophysics, University of Toronto, Toronto ON, M5S 3H4, Canada}

\author[0000-0001-9649-4815]{Bur\c{c}in Mutlu-Pakdil}
\affil{Department of Physics and Astronomy, Dartmouth College, Hanover, NH 03755, USA}

\author[0000-0002-0956-7949]{Kristine Spekkens}
\affiliation{Department of Physics and Space Science, Royal Military College of Canada P.O. Box 17000, Station Forces Kingston, ON K7K 7B4, Canada}
\affiliation{Department of Physics, Engineering Physics and Astronomy, Queen’s University, Kingston, ON K7L 3N6, Canada}



\begin{abstract}

We present new follow-up observations of two ultra-diffuse galaxies (UDGs)
selected for their distorted morphologies and tidal features, suggestive of tidal influence. Using Hubble Space Telescope Advanced Camera for Surveys F555W and F814W imaging, we identify $8\pm2$ globular clusters (GCs) in KUG~0203-Dw1 and $6\pm2$ in KDG~013, abundances typical for normal dwarf galaxies of similar stellar mass. Jansky Very Large Array data reveal a clear \hi\ detection of KUG~0203-Dw1 with a gas mass estimate of $\log{M_{\hi\ }/M_{\odot}}\lesssim 7.4$ and evidence of active stripping by the host, while KDG~013 has no clear gas detection. The UDGs likely originated as normal dwarf galaxies that have been subjected to significant stripping and tidal heating, causing them to become more diffuse. These two UDGs complete a sample of five exhibiting tidal features in the full Canada-France-Hawaii Telescope Legacy Survey area (CFHTLS; $\sim150^{2}$ deg). These tidally influenced UDGs exhibit diversse properties; one stands out as a potential result of a dwarf merger, while the remainder suggest tidal heating origins. We also cannot conclusively rule out that these galaxies became UDGs in the field before processing by the group environment, underscoring the need for broader searches of diffuse galaxies to better understand the impact of galaxy interactions.

\end{abstract}



\section{Introduction} 
\label{sec:intro}

Ultra diffuse galaxies (UDGs) lie at the extreme of galaxy evolution, given their unique combination of large size (half-light radii $>1.5$ kpc), extreme low surface brightness (central $g$-band surface brightness $>24$ mag arcsec$^{-2}$; \citealt{vandokkum2015,mihos2015}), and prevalence across all types of environments including clusters \citep{koda2015,mihos2015,vandokkum2015,yagi2016,venhola2017,wittmann2017,amorisco2018,lim2018,forbes2020,lim2020}, groups \citep{merritt2016,bennet2017,roman2017,vanderburg2017,spekkens2018,jones2021}, and the field \citep{martinez2016,leisman2017,greco2018,janowiecki2019,roman2019,prole2019,prole2021,jones2023}.  


The population of currently known UDGs overlaps with and extends previous LSB galaxy samples \citep[e.g.,][]{jerjen2000,conselice2002,conselice2003,mieske2007,deRijcke2009,penny2009,penny2011}.
However, the formation mechanisms of UDGs remain enigmatic, with proposed mechanisms including stellar feedback \citep{dicintio2017,chan2018}, high-spin host dark matter (DM) halos \citep{amorisco2016,rong2017}, early mergers \citep{wright2021,fielder2023}, failed galaxies \citep{vandokkum2016}, tidal heating of normal dwarf galaxies \citep{conselice2018,bennet2018,carleton2019,tremmel2020,jones2021}, and tidal dwarf galaxies (TDGs; \citealt{duc1998,duc2012,roman2021}). The diverse properties within the UDG population (e.g., gas content, UV emission, GC abundances, stellar masses, etc) suggest multiple formation pathways, or combinations thereof \citep[e.g.,][]{pandya2018,ruiz-lara2019,trujillo2019}. By focusing on UDGs with specific features or in particular environments, we can begin to disentangle formation pathways more clearly. In this study, we concentrate on a sample of UDGs exhibiting clear tidal features.

Various galaxy interactions can plausibly lead to the formation of UDGs. Early mergers of low-mass galaxies in hydrodynamical simulations result in present-day UDGs due to increased radius and angular momentum, leading to reduced central surface brightness \citep{wright2021}. Interactions involving massive, gas-rich galaxies may produce diffuse TDGs (\citealt{duc1998,duc2012,bennet2018,roman2021}). Tidal interactions arising from gravitational interplay between a low mass galaxy and a more massive neighbor (e.g., tidal heating) or entry into a cluster environment (e.g., ram pressure stripping and galaxy harassment) can redistribute stars, creating extended and low surface brightness features \citep[][]{roman2017,conselice2018,carleton2019,tremmel2020,jones2021}.

Key observational constraints that serve as discriminators for UDG formation pathways include their star cluster populations and \hi\ gas mass and morphology. Globular clusters (GCs) serve as cosmic fossils, tracing early star formation epochs \citep{renaud2018}. By characterizing their abundance and spatial distribution in a UDG we can infer essential details about its assembly history. For example, UDGs containing smaller GC populations are thought to have originated as dwarf galaxies in low mass DM halos, while UDGs with no GC population may be consistent with TDG formation. 
In tandem, the \hi\ in a galaxy is typically one of its most loosely bound baryonic components, making it a sensitive tracer of tidal interactions \citep[e.g.,][]{duc2013}. Therefore the morphology of the \hi\ within a UDG or even the absence of such a neutral gas reservoir can offer valuable insight into the evolutionary history of the system. 

Even with GC and \hi\ constraints, discerning UDG formation pathways in groups remains challenging. A study by \citet{jones2021} of two UDGs with tidal features (a subset of our sample) suggested they are likely puffed-up dwarfs due to the absence of \hi\ and typical GC abundances for dwarfs. However, \citet{jones2023} studied \hi\ bearing field UDGs that revealed a scarcity of globular clusters, aligning with expectations for similar-mass dwarfs. This result suggests that certain group UDGs may have been pre-existing entities accreted from the field rather than being transformed within the group environment (consistent with hydrodynamical simulation predictions, e.g., \citealt{jiang2019,liao2019}). \citet{fielder2023} studied a dramatically different UDG (another of our sample), with features pointing to a dwarf merger origin, indicating diverse formation pathways for UDGs within group environments. Collectively, these findings point to at least three distinct mechanisms contributing to UDG formation in group settings.

Here we present new data on two UDGs (KUG~0203-Dw1 and KDG~013) that contain evidence of a tidal disturbance, clear detections in the UV, and that lie in close proximity to a plausible group `host' galaxy. Utilizing Hubble Space Telescope (HST) and Jansky Very Large Array (VLA) observations, we investigate GC candidates, \hi\ mass, and morphology to distinguish between UDG formation scenarios for KUG~0203-Dw1 and KDG~013. 

These UDGs complete our sample of five total identified in a semi-automated search of the Canada–France–Hawaii Telescope Legacy Survey (CFHTLS). The search methodology is detailed in \citet{bennet2017}, with initial findings discussed in \citet{bennet2018}, where UDGs were identified that satisfy the surface brightness, size, and magnitude criteria of \citet{vandokkum2015} (see Table 1). This UDG sample can be split into those with associated UV emission and those without. The UDGs lacking UV were presented in \citet[][hereafter \citetalias{bennet2018}]{bennet2018} and \citet[][hereafter \citetalias{jones2021}]{jones2021}. The first UDG with a clear UV detection was presented in \citet[][hereafter \citetalias{fielder2023}]{fielder2023}. The two UDGs presented here also have clear UV detections. With this sample of five we can begin to assess the variety of properties exhibited by UDGs displaying tidal features, and determine the most plausible formation mechanisms of this sub-type of UDGs.

This manuscript is structured in the following way. \autoref{sec:hosts} briefly describes the environment of the two new UDG systems we present here. In \autoref{sec:observations} we describe the HST and VLA  data obtained for these two systems, along with additional supplementary observations employed to enhance the constraints on the properties of KUG~0203-Dw1 and KDG~013. In \autoref{sec:phys_props} we discuss our calculations and provide measurements of the physical properties of the two new UDGs. \autoref{sec:gcs} explores their GC systems. In \autoref{sec:formation} we discuss the formation mechanism for KUG~0203-Dw1 and KDG~013, and in \autoref{sec:comparison} we compare and contrast our full sample of five UDGs with tidal features identified in CFHTLS. We summarize and conclude in \autoref{sec:conclusion}. In \autoref{sec:appendix} we provide tables of the properties of GC candidates identified in KUG~0203-Dw1 and KDG~013 and additional supplementary figures.

\section{The UDG Host Systems}
\label{sec:hosts}

\begin{table*}[]
\centering
    \caption{Properties of Presumed UDG Hosts}
    \hspace{-0.6in}
    \begin{tabular}{rccccc}
    \hline\hline
    UDG & Galaxy & Morphology & $v_{\rm{helio}}$ & D & $M_{V}$ \\
    & & $\mathrm{km\,s^{-1}}$ & Mpc & mag \\ \hline
    KUG~0203-Dw1 & KUG~0203-100 & SB(s)m pec & $1885 \pm 1$ & $26.8 \pm 1.9$ & $\simeq-18.2$ \\ \hline
    KDG~013 & NGC~830 & SB0 & $3856 \pm 2$ & $55.5\pm3.9$ & $\simeq-20.7$ \\ 
     & NGC~829 & SB(s)c pec & $4074 \pm 2$ & $58.73\pm4.11$ & $\simeq-15.5$ \\ \hline
    \end{tabular} 
    \\[4pt]
    Columns: 3) Heliocentric velocity. For KUG~0203-100 this value comes from \citet{filho2013}. For NCG~830 and NGC~829 this value comes from \citet{sdssdr13}. 4) Virgo-infall distance. This value is derived from \citet{Karachentsev2004} for KUG~0203-100 and from \citet{sdssdr13} for NGC~830 and NGC~829. 5) $V$-band absolute magnitudes derived using
    SDSS magnitudes \citep[][]{sdssdr6} converted from AB to Vega and then converted to $V$ with documented SDSS conversions\tablenotemark{a}.\\
    \tablenotetext{a}{\url{http://classic.sdss.org/dr4/algorithms/sdssUBVRITransform.html}}
    \label{tab:hosts}
\end{table*}

Both of the UDGs that we present new data for in this work, KUG~0203-Dw1 and KDG~013, have associated galaxies in their respective groups which we briefly describe below.  Further measurements are provided in \autoref{tab:hosts}.\\
\textbf{KUG~0203-100:} This galaxy is the primary group member, with KUG~0203-Dw1 as the closest companion in addition to 3 other small galaxies. KUG~0203-100 (also known as SDSS J0205-0949) is an extremely metal poor galaxy, identified in \citet{kniazev2003}. 
\citet{pustilnik2007} notes that it exhibits a warped disk at both edges, a feature attributed to tidal interactions, and is rather luminous for such a low metallicity galaxy. 

\citet{pustilnik2007} also suggest that a nearby low surface brightness companion we now identify as KUG~0203-Dw1 may play a hand in the disturbed morphology of KUG~0203-100, which we expand upon in \autoref{sec:formation}. KUG~0203-Dw1 lies directly to the south-west of KUG~0203-100, with faint stellar material falling between the two (see \autoref{fig:tidal_features} and \autoref{fig:HI_img}; left columns). 
We adopt the Virgo in-fall distance of KUG~0203-100 for KUG~0203-Dw1.\\
\textbf{NGC~830:} The NGC~830 group contains NGC~829 and approximately 5 additional members 
of which these two are the dominant members. Their properties are summarized in \autoref{tab:hosts}. The heliocentric velocities of the other group members range from $3756\pm3$~$\kms$ to $4045\pm2$~$\kms$. In general many of the objects in the system are marginally disturbed, with a deformed disk present in NGC~829 and a stellar stream extending from the southwest of Mrk~1019. 

KDG~013 was first identified in \citet{karachentseva1968}, and was also identified in HYPERLEDA \citep{paturel2003} and the southern SMUDGes catalog (Systematically Measuring Ultra-diffuse Galaxies; \citealt{zaritsky2022}). It lies in the south-west part of the NGC~830 group, with faint stellar material extending predominantly in the northeast southwest directions but not directly connected to any other galaxy within the group (see \autoref{fig:HI_img} right column). For KDG~013 we adopt the distance of NGC~830 which is the approximate average distance of the other group members, ranging from $54.1\pm3.6$ to $58.7\pm4.1$~Mpc. However, despite it's morphology, it is unclear which galaxy in the group the UDG has interacted with.

\section{Observations}
\label{sec:observations}

\begin{figure*}
    \centering
    \includegraphics[width=0.98\textwidth]{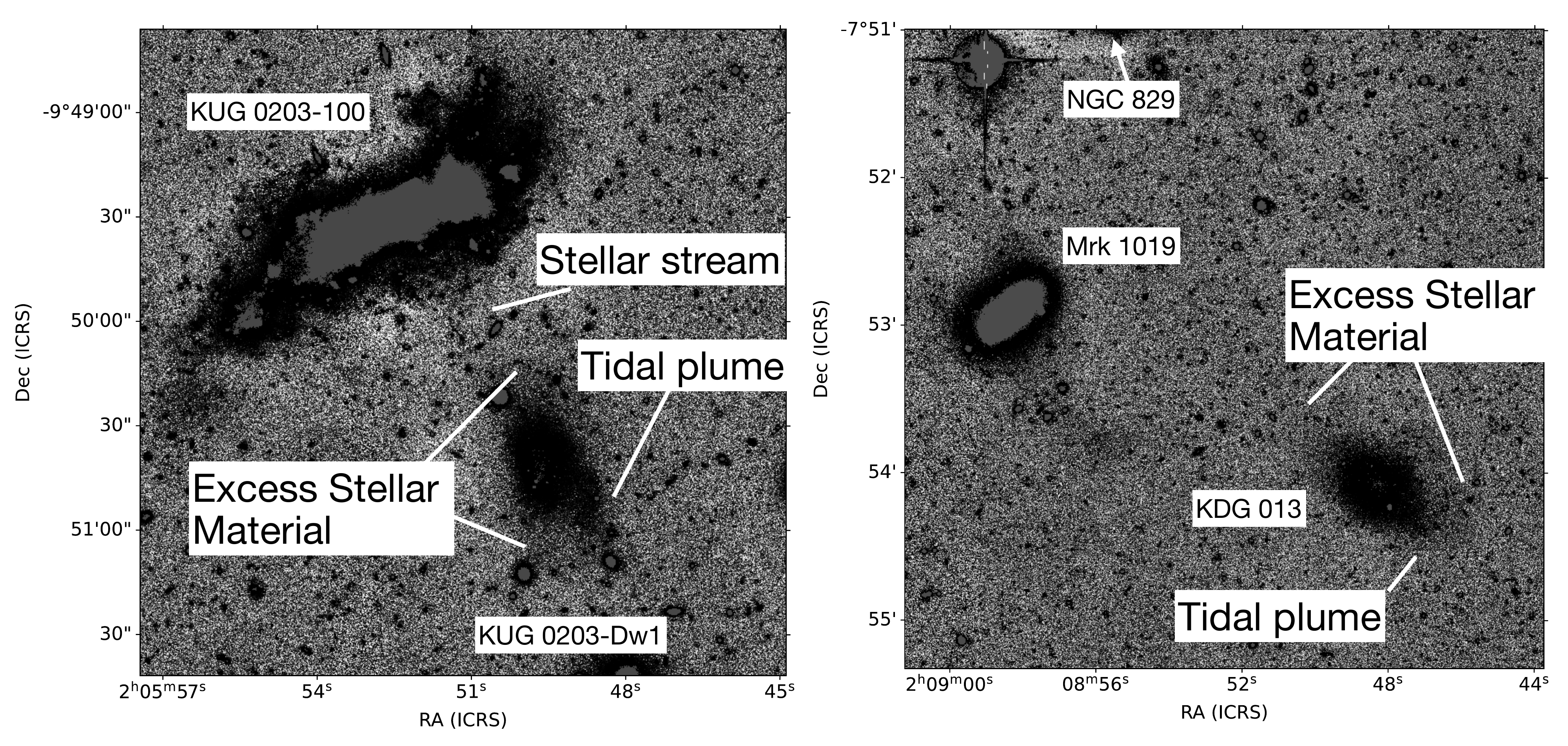}
    \caption{CFHTLS $r$-band imaging for KUG~0203-Dw1 (left) and KDG~013 (right), masked and scaled to emphasize low surface brightness features. There is a faint stellar stream between KUG~0203-Dw1 and KUG~0203-100. Additionally KUG~0203-Dw1 has a tidal plume that extends toward the southwest of it's main body, and extended stellar material northwest and south along the direction of the host. KDG~013 contains similar features compared to KUG~0203-Dw1, but with a smaller tidal plume. 
    The extended stellar material associated with KDG~013 falls along the direction of the majority of mass in the group (NGC~830, NGC~829, Mrk~1019).}
    \label{fig:tidal_features}
\end{figure*}

All of the UDGs studied in this work were originally identified in a semi-automated search for diffuse dwarfs in CFHTLS imaging \citep{bennet2017}. The initial findings are presented in \citetalias{bennet2018}, with the completed survey covering $\sim150\;\rm{deg}^{2}$. From the investigation area, five UDGs were selected for follow-up due to evidence of tidal disturbance through either possible associated tidal streams or elongated/amorphous morphology. High contrast imaging of KUG~0203-Dw1 and KDG~013 are presented in \autoref{fig:tidal_features} highlighting their respective features. 

\subsection{VLA Observations}

\begin{table*}[]
\centering
    \caption{Properties of VLA observations}
    \hspace{-1in}
    \begin{tabular}{cccccccc}
    \hline\hline
    Object & VLA config. & Bandwidth & \# channels & Band center             & Robust & Beam size & $\sigma_\mathrm{rms}$\tablenotemark{b} \\
    & & MHz & & $\mathrm{km\,s^{-1}}$ & & \arcsec$\times$\arcsec  & mJy/beam \\ \hline
    \multirow{3}{*}{KUG~203-Dw1} & C & 4 & 1024 & 1885 & 0.5 & $19\times15$ & 0.85 \\
    & D & 4 & 1024 & 1885 & 2.0 & $72\times56$ & 0.98 \\
    & C+D & 4 & 1024 & 1885 & 0.5 & $27\times22$ & 0.47 \\ \hline 
    KDG~013-Dw1 & D & 8 & 2048 & 4074 & 2.0 & $74\times60$ & 0.77 \\ \hline
    \end{tabular}
    \tablenotetext{b}{RMS noise in 5~\kms \ channels.}
    \label{tab:VLAobs}
\end{table*}

\begin{figure*}
    \centering
    \includegraphics[width=1.04\columnwidth]{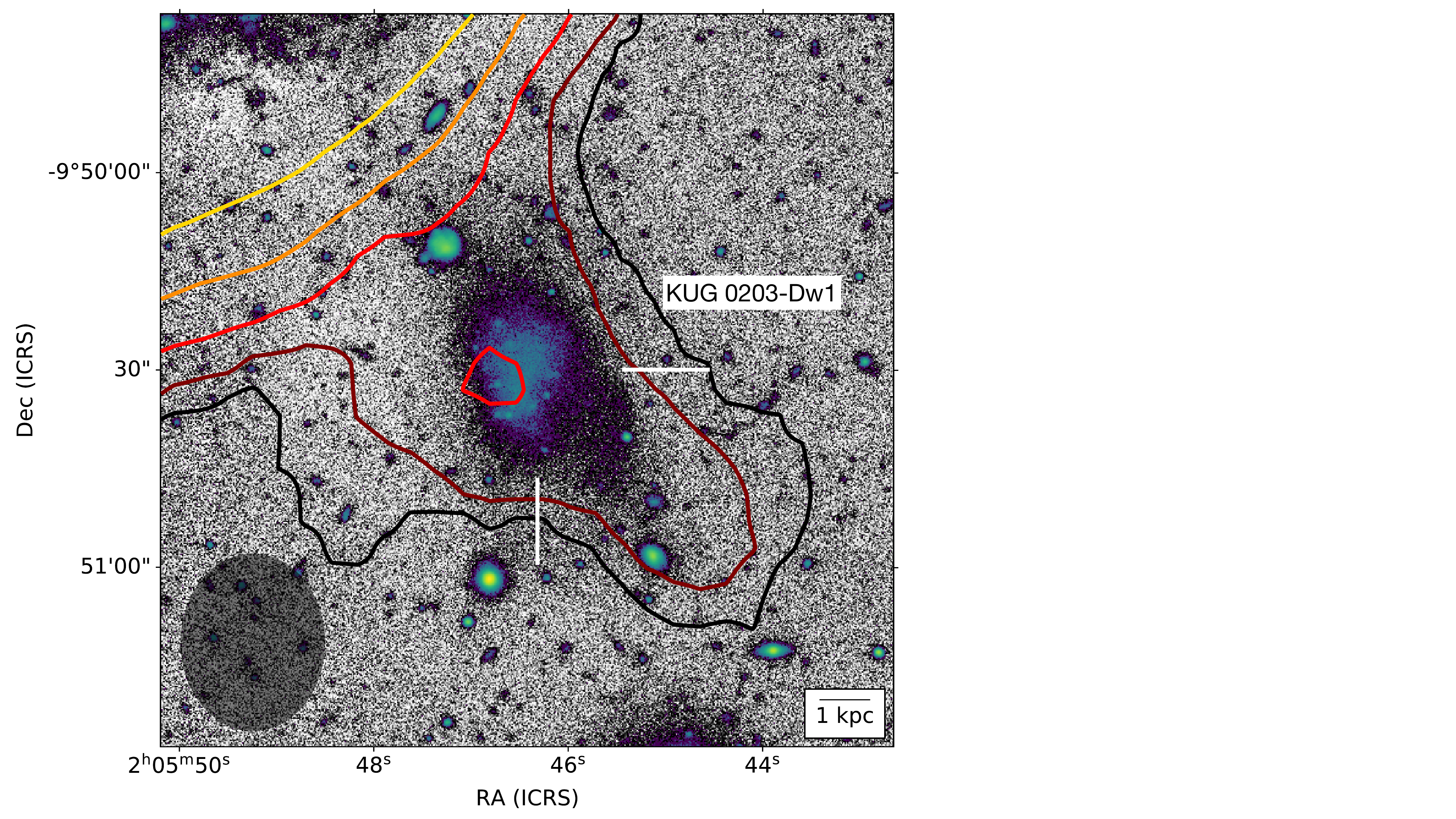}
    \includegraphics[width=\columnwidth]{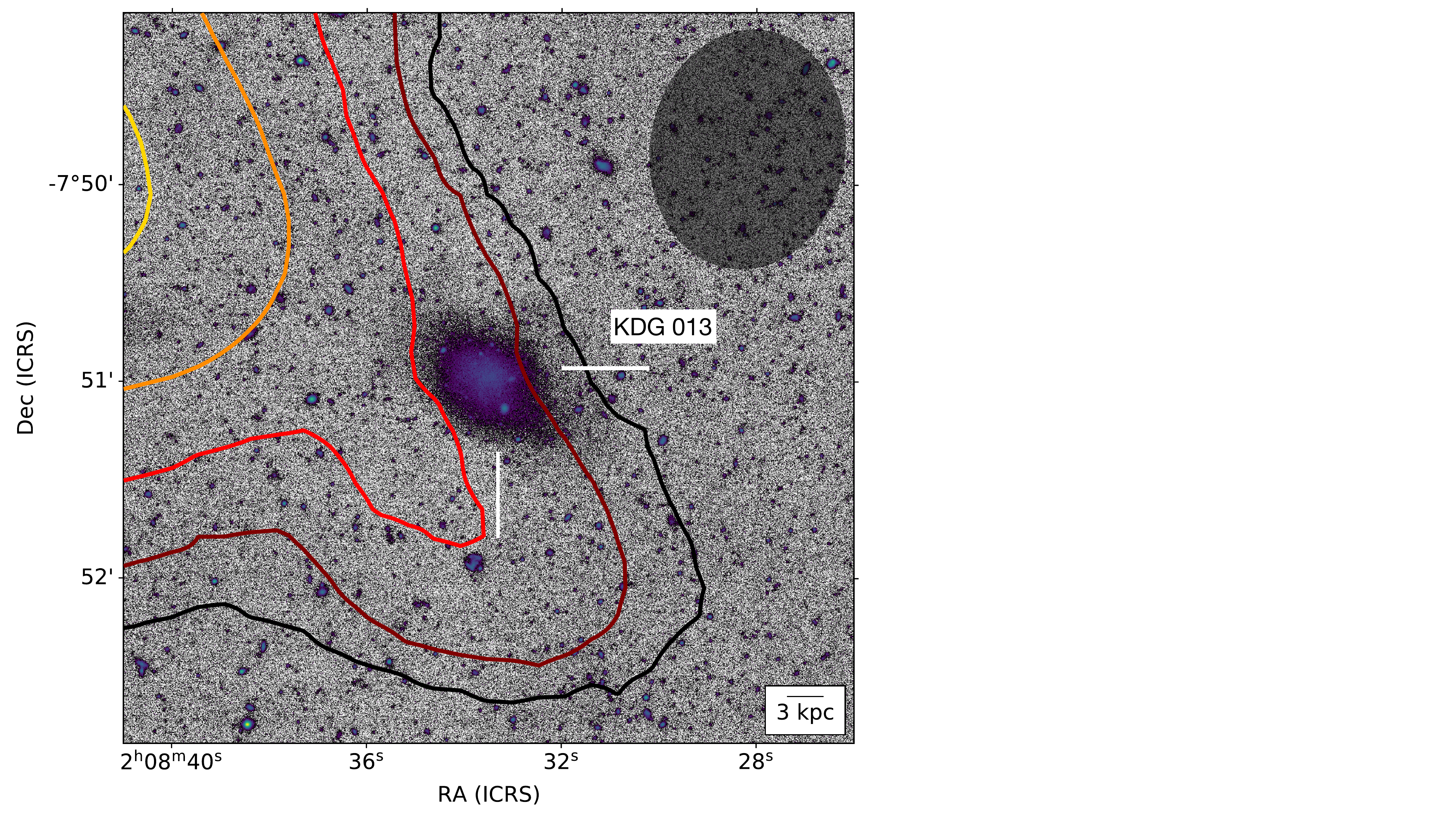} \\
    \includegraphics[width=\columnwidth]{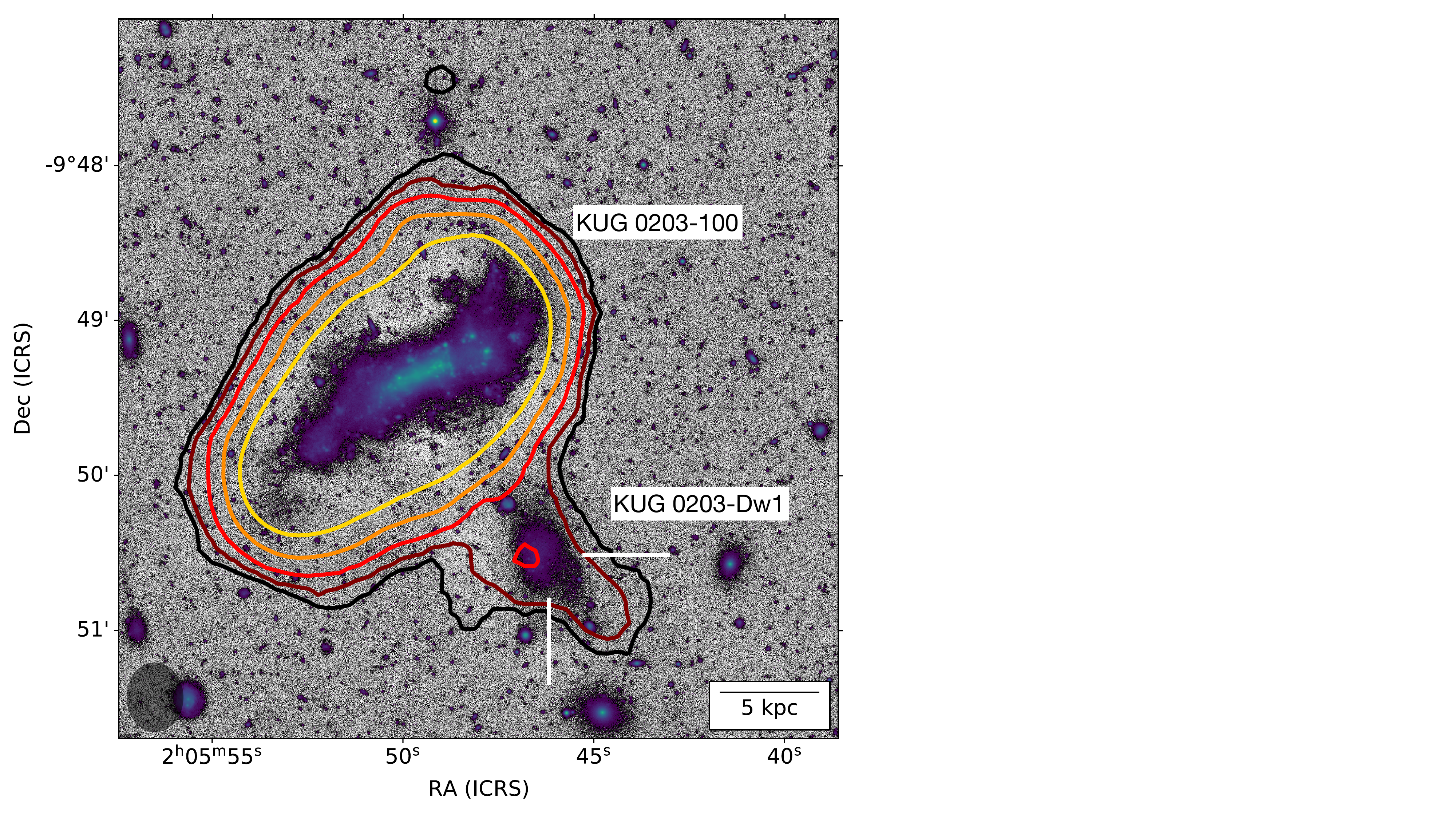}
    \includegraphics[width=\columnwidth]{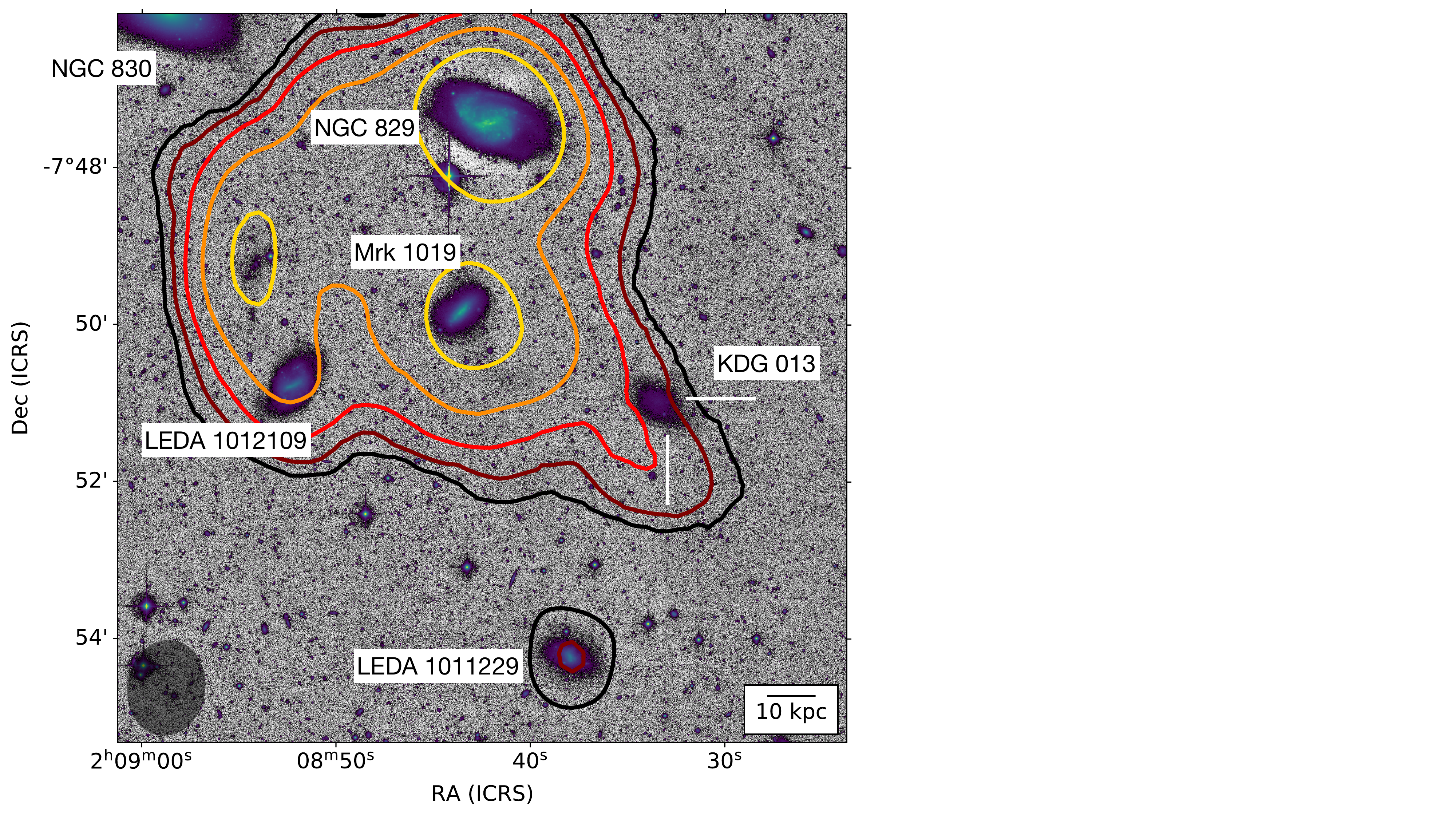}
    \caption{VLA \hi\ contours overlaid on CFHTLS imaging for KUG~0203-Dw1 (left column) and KDG~013 (right column). The top row is centered on the UDG while the bottom row shows a wider view of the respective galaxy group. \textit{Upper left:} A high-contrast CFHT $g$-band image of KUG~0203-Dw1 with integrated VLA C+D array \hi\ emission contours overlaid. The minimum contour is 
    $1.3\times10^{19} \; \mathrm{cm^{-2}}$ ($3\sigma$) in column density (over 10~\kms). Each contour is double the previous. The synthesized beam is shown as a semitransparent ellipse in the lower left corner, which corresponds to a beam size of $27.0 \times 22.0$ arcsec$^2$.
    \textit{Lower left:} Same as upper left, but showing the full system with KUG~0203-100. A small over-density to the north of KUG~0203-100 (black contout) is a real detection and not noise. \textit{Upper right:} Similar to the left panels, for KDG~013. The \hi\ emission is derived from VLA D-array only. Here the minimum contour is $8.4\times10^{18} \; \mathrm{cm^{-2}}$ ($3\sigma$) in column density (over 20~\kms). The synthesized beam size corresponds to $73.5 \times 59.6$ arcsec$^2$. \textit{Lower right:} Same as upper right, but showing a majority of the NGC~830 group. }
    \label{fig:HI_img}
\end{figure*}

KUG~0203-Dw1 and KDG~013 were observed in the VLA D-configuration during July 2022, as a part of project 22A-225 (PI: D.~Sand). Each target had a total on-source integration time of approximately 4~h. Both targets were observed with a velocity resolution of $\sim$0.8~$\kms$ (averaged to 5~$\kms$ during imaging). A 4~MHz band (1024 channels) was used for KUG~0203-Dw1 and an 8~MHz band (2048 channels) for KDG~013. These bands were centered at 1885~\kms \ and 4074~\kms \ for KUG~0203-Dw1 and KDG~013, respectively.\footnote{In addition, the data were also recorded simultaneously at a lower spectral resolution (62.5~kHz per channel, or $\sim$13~\kms) over a bandwidth of 32~MHz, providing broad coverage in velocity ($\sim$7000~\kms). However, in practice no significant \hi \ signal at the target locations was detected beyond the narrower, higher spectral resolution band described in the main text, and we do not consider the low resolution data further here.} KUG~0203-Dw1 was also selected for follow-up under Director's Discretionary Time in the VLA C-configuration during winter of 2022 (PI: C.~Fielder), with the same spectral configuration and a total on source integration time of 4~h. The results of these observations are presented in \autoref{fig:HI_img}. For KUG~0203-Dw1 we imaged the C and D array data both separately and together, with the latter used for the contour plots (referred to as VLA C+D data). The properties of these observations are summarized in Table~\ref{tab:VLAobs}.

The data were reduced with the \hi\ pipeline\footnote{\url{https://github.com/AMIGA-IAA/hcg_hi_pipeline}} of \citet{jones2023b}. For a full description we refer the reader to that work. In brief the pipeline performs both manual and automated flagging before proceeding with gain and phase calibrations, $UV$ continuum subtraction, and imaging using standard \texttt{CASA} tasks (Common Astronomy Software Applications; \citealt{mcmullin2007}). 

The KUG~0203-Dw1 data suffered from moderate interference in C-configuration where $\sim$19\% of visibilities were flagged, while $\sim$42\% were flagged in D-configuration. The higher fraction of flagged data in D-configuration is mostly the result of four antenna failures.
A Briggs robust parameter of 0.5 was used during imaging for the C+D data to offer a compromise between sensitivity and resolution. The synthesized beam size was determined to be $27.0\times22.0$ arcsec$^2$ and the central velocity is 1885~km s$^{-1}$ (a similar measurement to the host).

The KDG~013 D-array data had moderate interference with $\sim$22\% of visibilities flagged. 
Imaging was conducted with automated masking, applying a Briggs robust parameter of 2.0 (i.e. natural weighting) in order to maximize sensitivity to the extended features. This resulted in a synthesized beam size of $74\times60$ arcsec$^2$.

For the generation of the VLA moment zero maps in \autoref{fig:HI_img}, we utilized the \texttt{SoFiA} software \citep{serra2015} to create a source mask. Employing the standard smooth and clip algorithm, we conducted the process both with no smoothing and with Gaussian smoothing kernels approximately 0.5 and 1.0 times the beam size. Additionally, we applied boxcar spectral smoothing over 0 and 3 channels. The clip threshold was set at $3.5\sigma$, and to ensure robustness, we imposed a stringent 95\% reliability threshold.

\subsection{HST Observations}

KUG~0203-Dw1 and KDG~013 were observed in July and September of 2022, under HST program ID 16890 \citep{sand2021}. Both targets were observed by the Advanced Camera for Surveys (ACS) with the Wide Field Channel (WFC) in the F555W and F814W filters. The total integration time for KUG~0203-Dw1 is 2046~s in F555W and 2098~s in F814W. Integration times for KDG~013 correspond to 2046~s and 2089~s in F555W and F814W respectively. In addition, Wide Field Camera 3 (WFC3) images were taken in parallel to utilize as reference background fields as needed. 

\begin{figure*}
    \centering
    \includegraphics[width=\columnwidth]{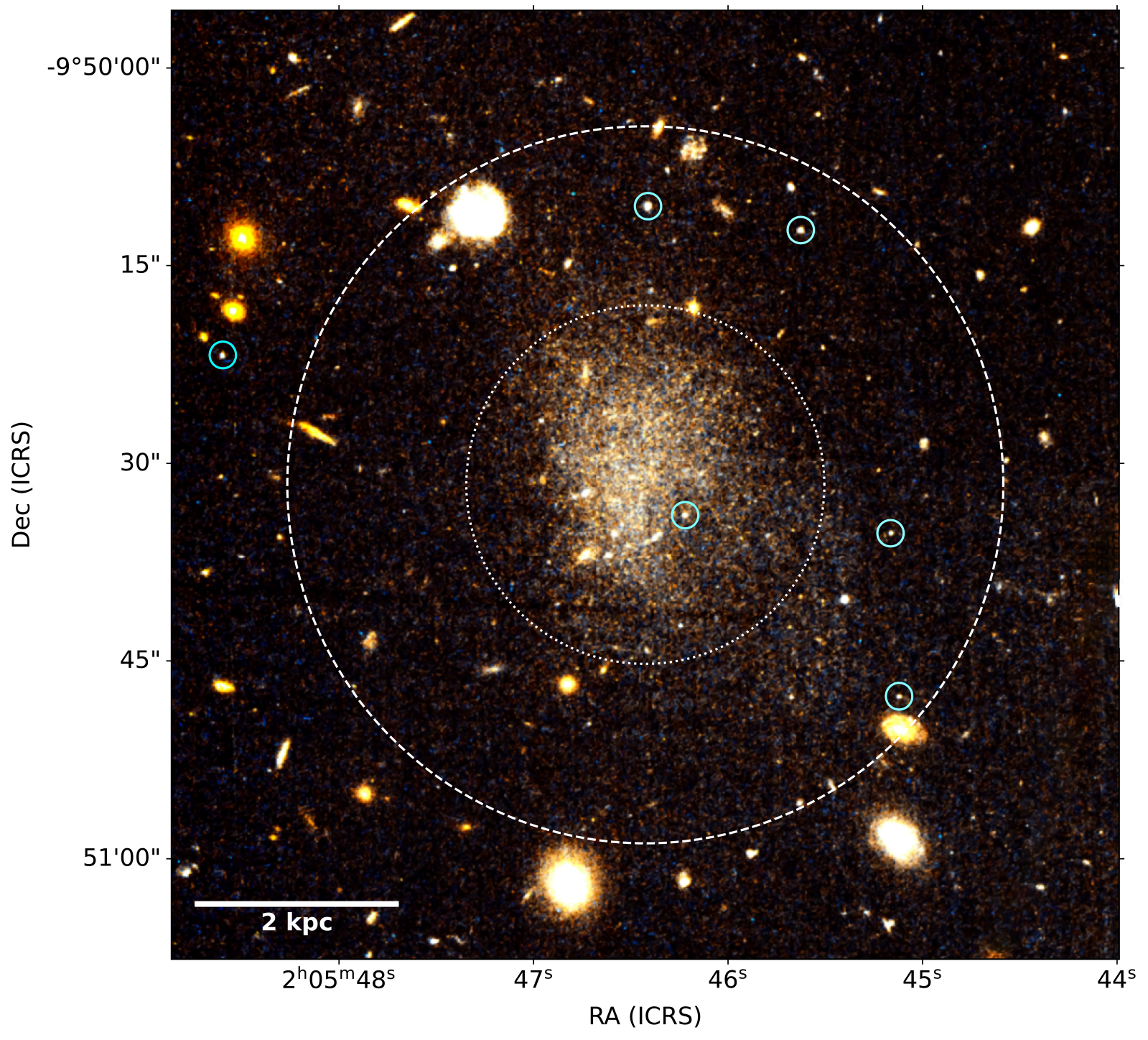}
    \includegraphics[width=\columnwidth]{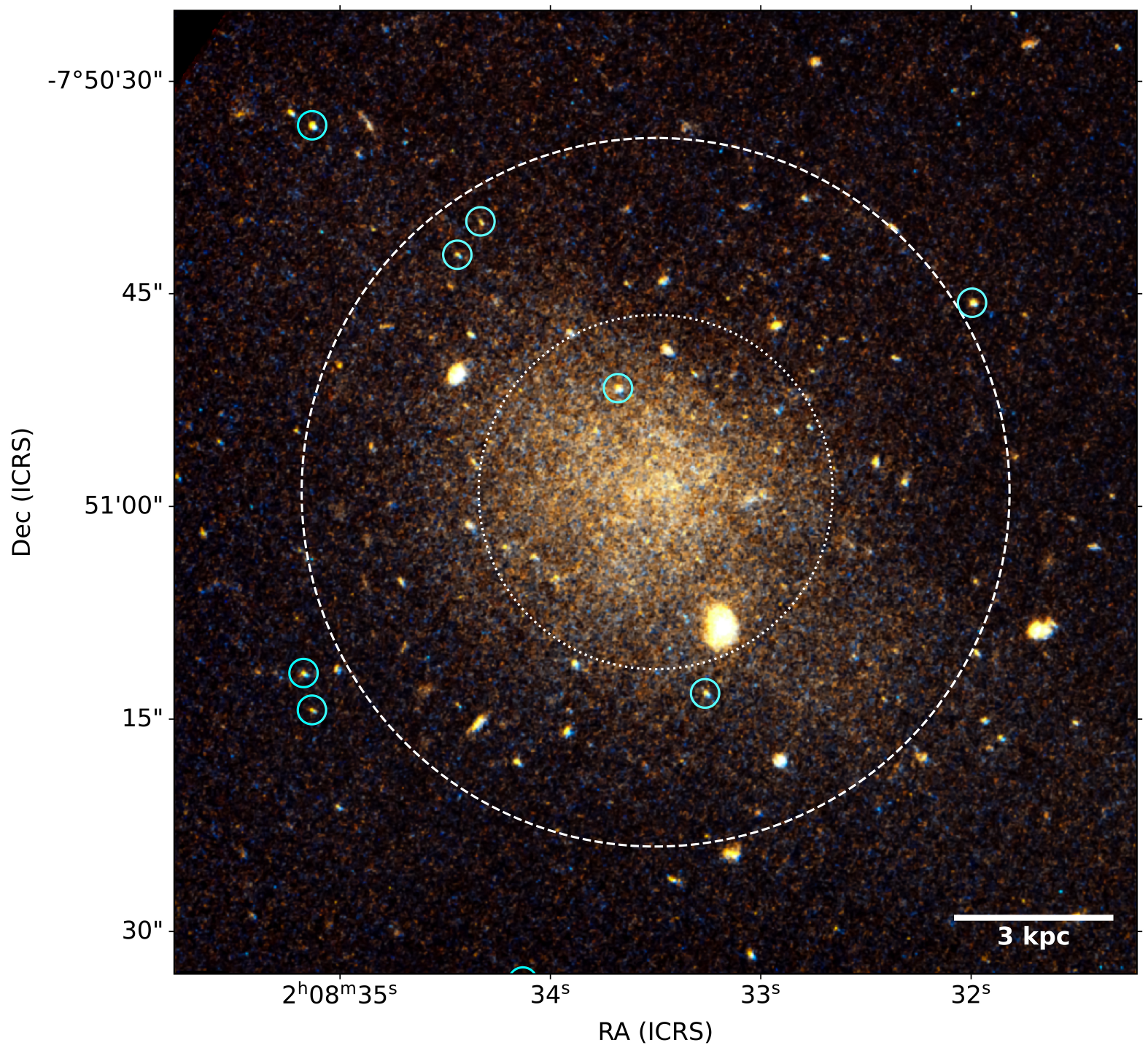}\\
    \includegraphics[width=\columnwidth]{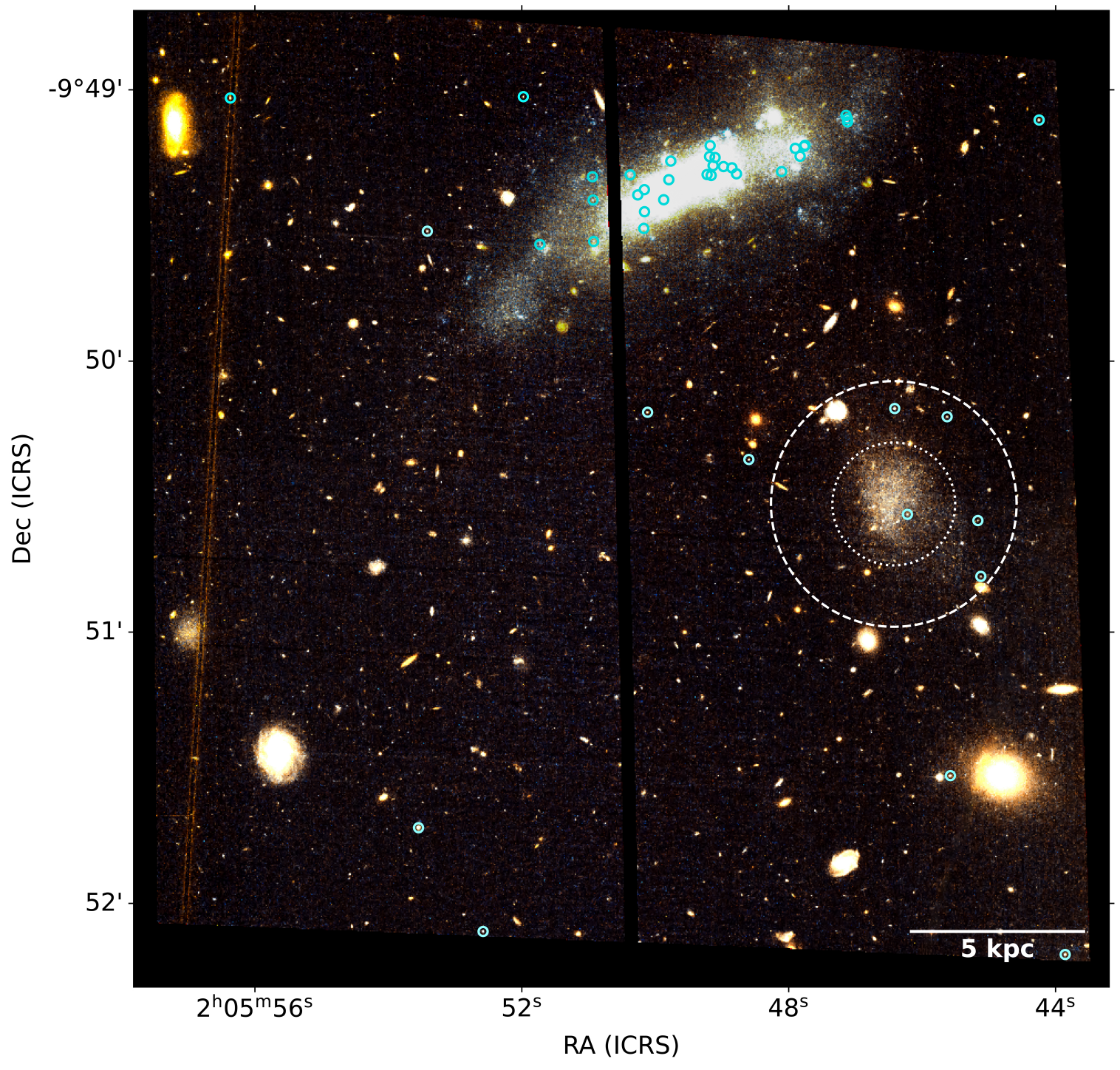}
    \includegraphics[width=\columnwidth]{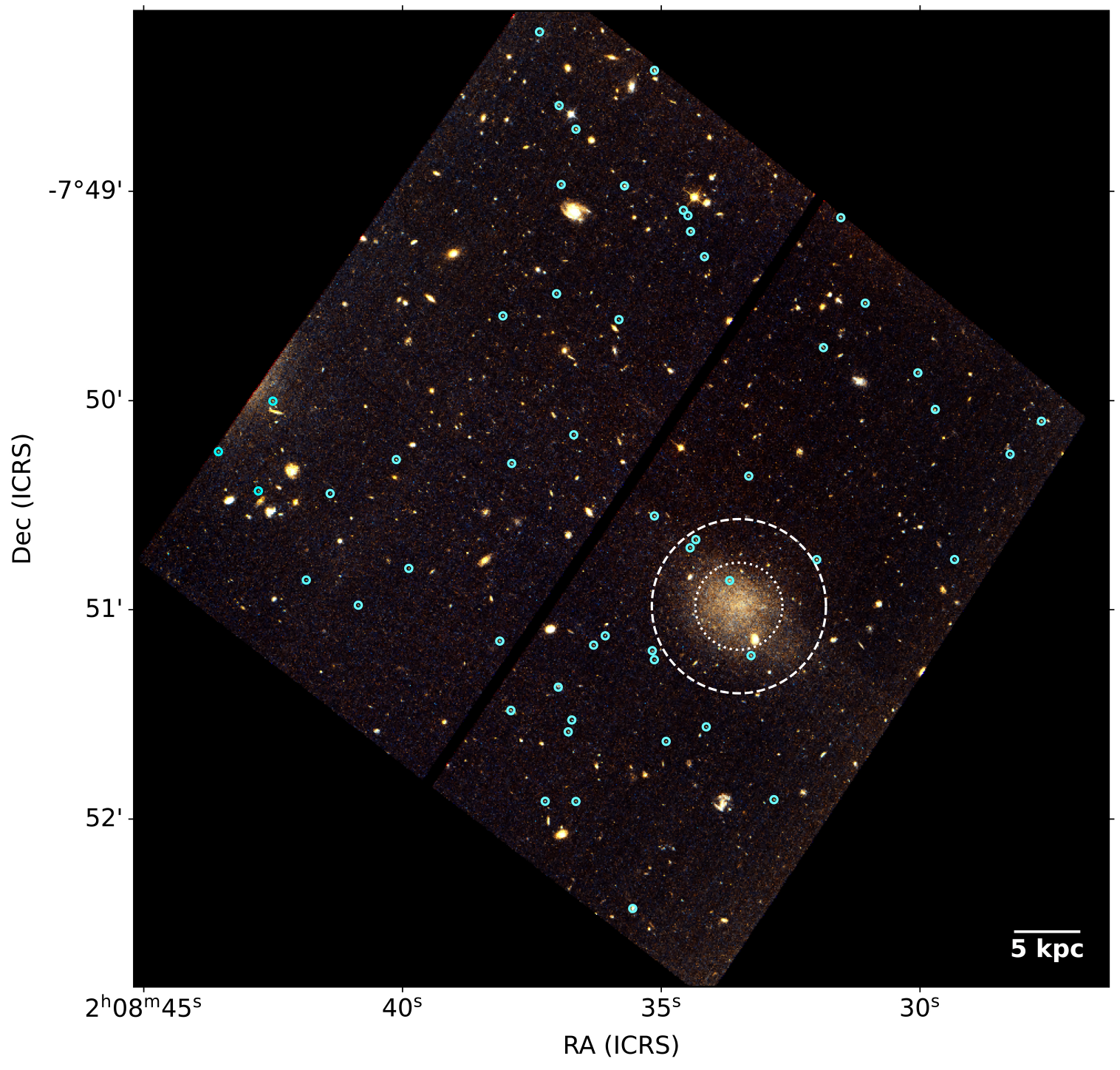}
    \caption{HST ACS/WFC F555W/F814W composite color images of KUG~0203-Dw1 (left column) and KDG~013 (right column). The upper row is centered on the UDG while the lower row shows the full HST frame. \textit{Upper Left}: HST ACS/WFC F555W/F814W composite color image of KUG~0203-Dw1. The inner, white dotted circular aperture marks $r_{1/2}$ (11.8''), and the dashed circular aperture marks $2\times r_{1/2}$ (23.6'') which we select as the outer edge of the UDG. Small cyan apertures enclose all GCCs within the image, with a total of 5 within the aperture of the UDG. 
    \textit{Lower Left}: The full HST image of KUG~0203-Dw1, which includes its host galaxy KUG~0203-100. 
    \textit{Upper Right}: Same as the left panel, but for KDG~013. Here the apertures correspond to 12.5'' and 25'', with a total of 4 GCCs. GCCs of either UDG are selected from within the $2\times r_{1/2}$ UDG boundary. 
    \textit{Lower right:} The full HST image of KDG~013. The edge of Mrk 1019 is visible to the northeast of KDG~013. 
    }
    \label{fig:hst_img}
\end{figure*}

We present RGB color composite images for KUG~0203-Dw1 and KDG~013 in \autoref{fig:hst_img}, constructed from the combined F555W and F814W exposures. Highlighted sources are discussed in detail in \autoref{sec:gcs}. After obtaining the standard data products from the STScI archive, the individual exposures were aligned and background subtracted using the \textsc{Dolphot 2.0} software \citep{dolphin2000, dolphin2016} with the standard ACS and WFC3 parameters provided in the user manual. Then, a combined point source catalog for each of the fields was generated. \textsc{Dolphot} used the HST to Johnson-Cousins magnitude conversion factors presented in \citet{sirianni2005} to determine the $V$- and $I$-band magnitudes from F555W and F815W magnitudes. The photometry was then corrected for Galactic extinction using the NASA/IPAC\footnote{\url{https://irsa.ipac.caltech.edu/applications/DUST/}} online tool, with values derived from the \citet{schlafly2011} extinction coefficients. Unless otherwise indicated, all magnitudes presented in this work are Milky Way extinction-corrected Vega magnitudes.

Completeness limits are determined via artificial star tests with the tools provided by \textsc{Dolphot}. We place nearly $100,000$ artificial stars into the ACS field, which span a large color range from $-$1 to 2 in F555W-F814W (well beyond the range used to select GCs). We then measure the fraction of those stars that we recover as a function of apparent magnitude. For KUG~0203-Dw1 we find that we are $90\%$ complete to $m_{F814W} = 26.0$ and $50\%$ complete to $m_{F814W} = 26.8$. For KDG~013 we are $90\%$ complete to $m_{F814W} = 26.3$ and $50\%$ complete to $m_{F814W} = 27.1$. Note that using the \citet{sirianni2005} conversions the completeness limits are the same in $I$-band for the $(V-I)$ color range of GCs.

\subsection{Supplementary Observations}

\subsubsection{CFHT}
The UDGs were first identified in the CFHTLS. Specifically, our observations are supplemented with the Wide portion of the survey, conducted between 2003 and 2009 and covering 171 square degrees in the $u$, $g$, $r$, $i$ \& $z$ bands. Using the nomenclature of Figure 4 of \citet{gwyn12}, KUG~0203-Dw1 was observed in the W$1-3-3$ tile and KDG~013 was observed in the W$1-2-1$ tile. Exposure times for each stack and filter varies, with 2500~s in $g$, 2180~s in $r$, and 4305~s in $i$ for the W$1-3-3$ tile. For the W$1-2-1$ tile the exposure times are 5000~s in $g$, 3180~s in $r$, and 4305~s in $i$. The tiles were downloaded directly from the Canadian Astronomy Data Centre (CADC) and have been processed by the Terapix 7 pipeline. The point spread functions (PSFs) for those image stacks were also downloaded from the CADC, which were used for measuring dwarf structural parameters. The construction and calibration of these utilized the MegaPipe data pipeline \citep{gwyn08} and is described in detail by \citet{gwyn12}. The CFHT images are used to derive the physical properties of the UDGs given the increased surface brightness sensitivity compared to HST ACS/WFC (see \citetalias{jones2021} for further discussion). High-contrast CFHT $g$-band images are shown in \autoref{fig:tidal_features} and utilized as the background for our radio contours in \autoref{fig:HI_img}.

\subsubsection{GALEX Ultraviolet Observations}

Data from GALEX \citep{galex2005} were used to measure the star formation rates of KUG~0203-Dw1 and KDG~013. Both targets were observed in NUV for $\sim$1600 s as part of the guest investigator program (GI5-028, PI: Balogh), which did not include FUV observations. KUG~0203-Dw1 was also observed as part of the GALEX all-sky survey at medium depth for $\sim$1600 s in both NUV and FUV. In \autoref{fig:UV_img} we provide the GALEX images of both UDGs. 

\section{Physical Properties}
\label{sec:phys_props}

The UDGs considered in this work contain evidence of past or ongoing interactions, manifest in tidal features and/or disturbed morphologies. These objects have large half-light radii ($>1.5$~kpc) and faint central surface brightnesses (central $g$-band surface brightness $>24$ mag arcsec$^{-2}$), meaning that they fit into the standard definition of a UDG (see \autoref{subsec:optical_props} for discussion on how these quantities are derived). The physical properties discussed in the following sub-sections are summarized in \autoref{tab:udg_props}. Note that the section to the right of the vertical divide in this table reproduces the literature values of the respective UDGs presented in \citetalias{fielder2023}, \citetalias{jones2021}, and \citetalias{bennet2018}. These quantities were all derived in the same manner as the UDGs presented in this work. 

The optical tidal morphology of both of these UDGs is highlighted in \autoref{fig:tidal_features}, with the \hi\ morphology displayed in \autoref{fig:HI_img}. KUG~0203-Dw1 exhibits elongation in the northeast-southwest direction, perpendicular to the disk of KUG~0203-100. This galaxy displays a somewhat clumpy appearance, accentuated by a discernible faint tidal plume evident in the high contrast CFHT image, extending from the southwest. A stellar stream also lies between KUG~0203-Dw1 and KUG~0203-100.
KDG~013 shares a comparable elongated morphology in the northeast-southwest direction, pointing towards several galaxies in the group (NGC~830, NGC~829, and Mrk~1019). There are more clear clumps in the southwest with additional diffuse material in the northwest (features more obvious in the high contrast CFHT images). There is also a small faint tidal plume similarly extending from the southwest.

\subsection{HI Properties}
\label{subsec:hi_props}

\subsubsection{KUG~0203-Dw1}

The \hi\ morphology of KUG~0203-Dw1 is complex, with a bridge connecting to KUG~0203-100 and a spur extending to the southeast. However, there is a clear concentration of \hi\ at the center of the UDG (left column, \autoref{fig:HI_img}). In the zoomed-out lower-left panel of \autoref{fig:HI_img} there is a small feature to the north of KDG~0203-100 (black contour) 
This is a real detection and not noise, as this feature is visible in the separate C-configuration and D-configuration data cubes as well as the combined C+D data. It is likely the end of a trailing/leading tail of \hi\ gas, where the rest is blended with the \hi\ emission of the host. 

The total mass of the HI feature that extends to the south of KUG~0203-100 is approximately $\log{M_{\hi\ }/M_{\odot}} = 7.6$. The mass that is gravitationally bound to KUG~0203-Dw1 is certainly less than this value. To obtain an approximate upper limit we placed an aperture equal in area to the synthesized beam at the location of the peak of \hi\ emission centered on KUG~0203-Dw1. This gave a value of $\log{M_{\hi\ }/M_{\odot}} = 7.4$. We note that this is not a strict upper limit as the beam is only slightly smaller than the entire feature, thus this approach may still include some emission from gas that is not bound to the UDG. However, this provides a reasonable approximation for the \hi\ mass of KUG~0203-Dw1.

\begin{figure*}[t]
    \centering
    \includegraphics[width=\columnwidth]{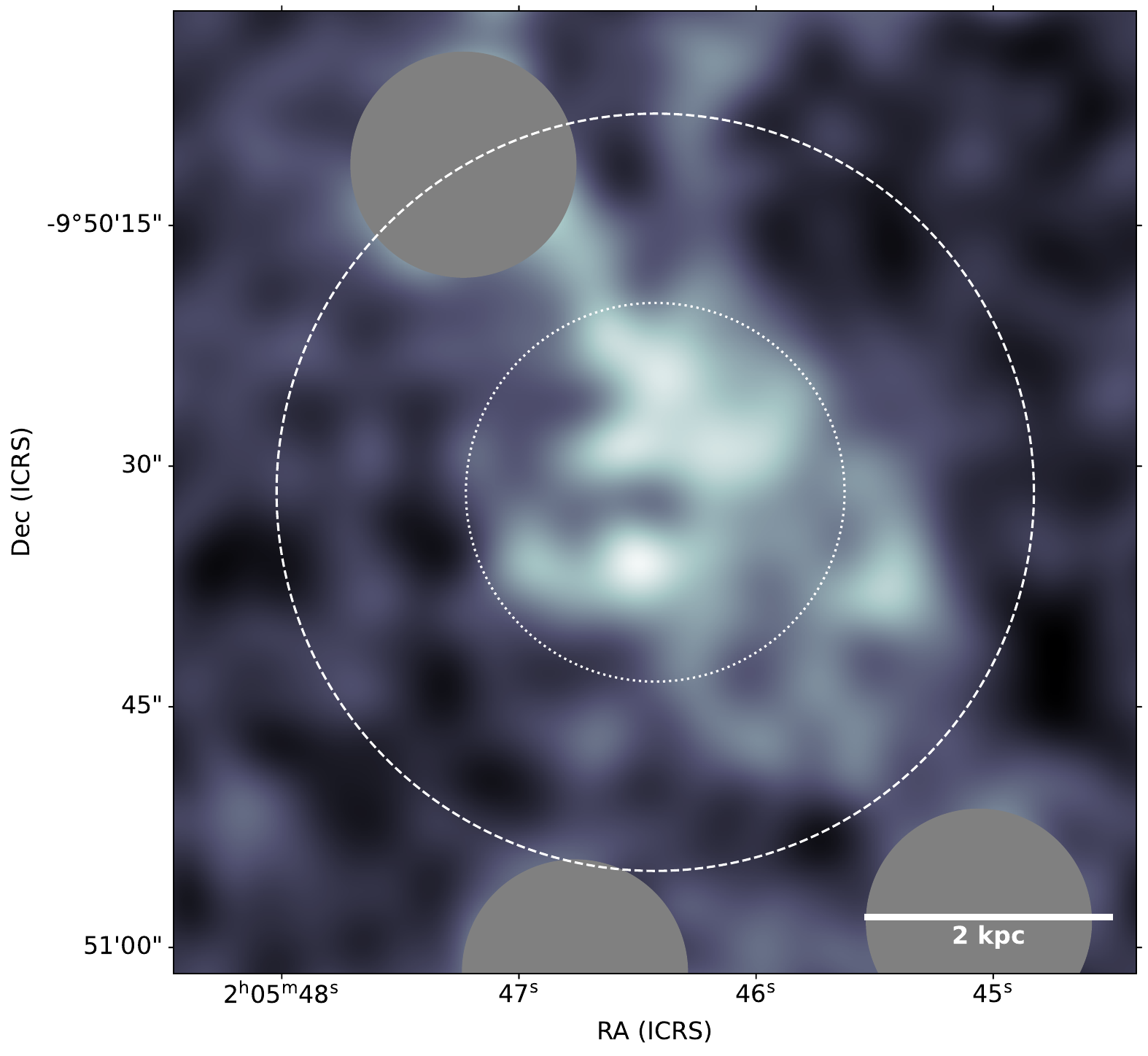}
    \includegraphics[width=0.99\columnwidth]{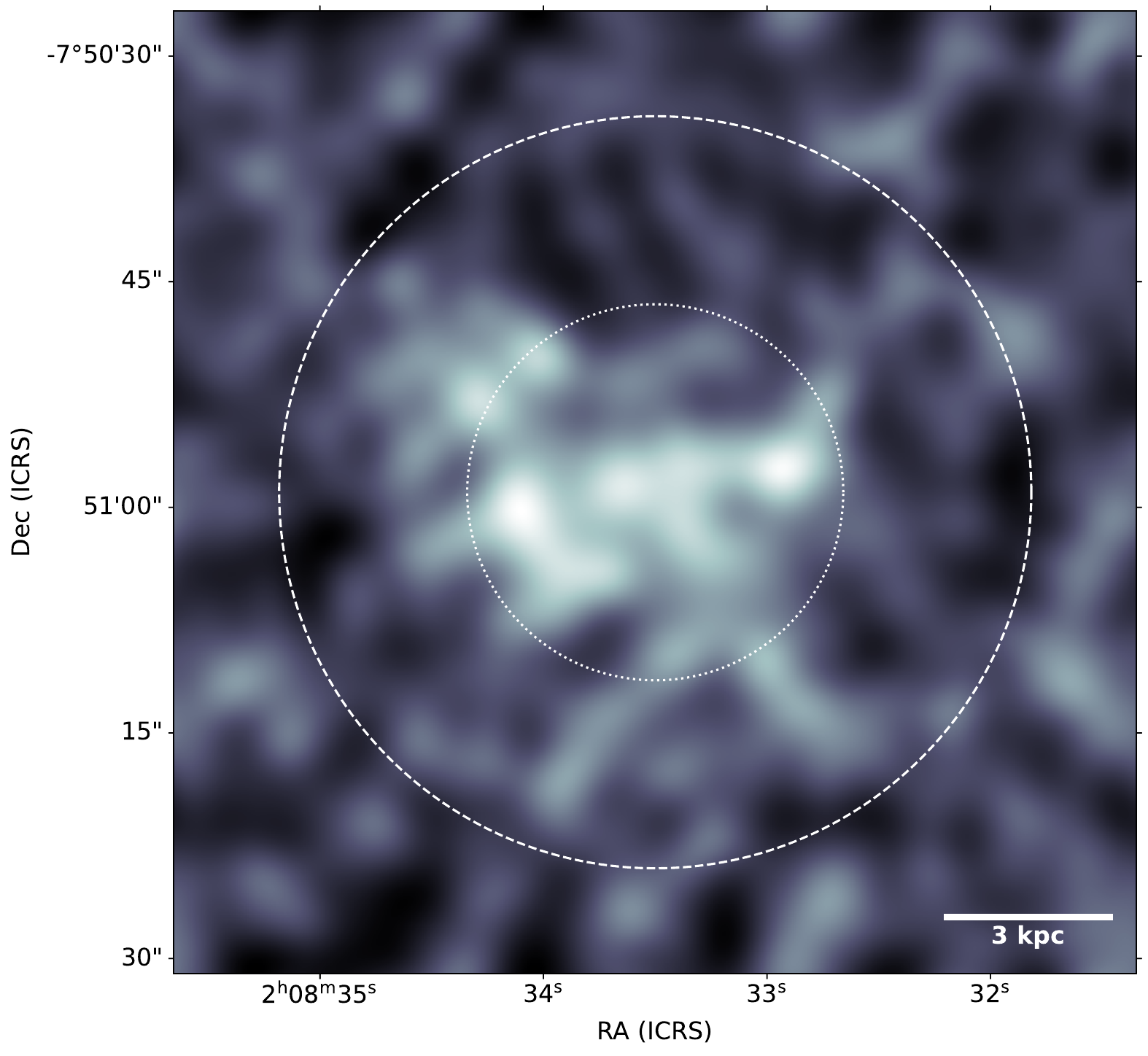}
    \caption{\textit{Left:} Composite GALEX FUV and NUV image of KUG~0203-Dw1. Three background galaxies fall very close to the UDG, so we apply a mask setting values above the threshold of $0.008$ to $0$. The positions of these objects are marked by gray circles. \textit{Right:} GALEX NUV image of KDG~013. Circular apertures are the same as those shown in \autoref{fig:hst_img}. No threshold was applied to this image as it was not necessary. Both images have the same scaling.}
    \label{fig:UV_img}
\end{figure*}

\subsubsection{KDG~013}

The \hi\ morphology of the NGC~830 group is large and complex. In the vicinity of KDG~013, four galaxies show \hi\ detections, with three embedded in a cloud of extended gas features. This includes a notable intergalactic concentration of gas and star formation just north of LEDA~1012109 (see the oval yellow contour in the lower right panel of \autoref{fig:HI_img}) which may be a source like those discovered in \citep{jones2022}. A spur of this complex \hi\ distribution extends towards KDG~013, overlapping with it at the resolution of the VLA observations. However, there is no evidence of an \hi\ concentration directly centered on the UDG. As such, it is improbable that KDG~013 harbors a significant gas reservoir. Similarly, it is unlikely that KDG~013 is related to the gas disturbance given that the feature does not align with KDG~013 (e.g., see the red contour in the right column of \autoref{fig:HI_img}), but a conclusive determination would require comparing the optical redshift of the UDG to that of the gas. Given the disorganized nature of the gas in the NGC~830 group, the gas overlapping with KDG~013 could originate from one of the other galaxies in the group (especially Mrk~1019 given its stellar stream), or may be a result of the gravitational influence of the UDG pulling the intra-group medium.

Based on the contours in the image and the beam size, if KDG~013 had an \hi\ mass above $\log{M_{\hi\ }/M_{\odot}} = 8.6$ then it would certainly cause a visible concentration in the contours. This places a weak upper limit on the \hi\ content of KDG~013, which is likely considerably lower. For instance, if KDG~013 is not close in velocity to the gas cloud it is projected near, the constraint on its \hi\ content would be the point source detection limit of our observations, which is $\log{M_{\hi\ }/M_{\odot}}= 8.1$ at $3\sigma$. We do not detect any other significant \hi \ signal at this location throughout the entire band of our VLA observation ($660-7560$~\kms), but note that the \hi \ mass limit is distance-dependent and assumes the host distance (55.5~Mpc).

\subsection{Optical/UV Properties}
\label{subsec:optical_props}

The optical/UV properties of both UDGs are determined by GALFIT \citep{peng2002}, using the same methodology presented in \citet{bennet2017,bennet2018} \citep[see also][]{merritt2016}. The optical properties are derived from the CFHT data, but the HST data has been shown to provide near identical results (see \citetalias{jones2021}).  
Both galaxies were modeled using a standard Sérsic profile \citep{sersic1963}, but KUG~0203-Dw1 was more of a challenge for GALFIT to constrain and ultimately required two different Sérsic profiles to fit. Otherwise all fit parameters are allowed to vary without restriction. The uncertainties on the parameters for each UDG are determined by injecting 100 simulated UDGs into the image, each with the best fitting properties of the given UDG as determined by GALFIT. Then GALFIT is used to recover the properties of the simulated UDGs. The scatter on the recovered properties of the simulated UDGs represents the uncertainty on the best fit properties of the real UDG. We present our parameter estimates in \autoref{tab:udg_props}. $FUV$, $NUV$, $g$, and $r$-band magnitudes are presented in the AB system, while $V$ and $I$-band magnitudes remain in the Vega system. CFHT $g$ and $r$-band data is converted to $V$ and $I$ following \citet{jester2005} (see also footnote $a$ of \autoref{tab:hosts}). The best-fit GALFIT models both have an ellipticity, consistent with the elongated morphology of the UDGs, as well as a circularized radius. Note that some of the CFHT filters suffered from over-subtraction. To mitigate this issue general properties like radius were measured in filters where this was not an issue and then fixed when determining other measurements from more problematic filters. Errors associated with the over-subtraction are significantly smaller than the errors reported.

Stellar masses are derived using the relations from Table 1 of \citet{zhang2017} and our CFHT derived $(V-I)$ and $M_{V}$. We derive a stellar mass-to-light ratio $\Gamma_{*}=0.5$ for KUG~0203-Dw1 and $\Gamma_{*}=0.9$ for KDG~013 which we then convert to stellar mass. We caution the reader to see this is an estimate from a model and not a derived quantity. 

The UV data of both UDGs originate from GALEX \citep{galex2005}. Foreground sources are first masked (see the left panel of \autoref{fig:UV_img}). Then UV fluxes are determined via aperture photometry, with an aperture radius equivalent to two half-light radii of the best fit GALFIT model to the optical data (see the dashed aperture in \autoref{fig:UV_img}). We derive star formation rates using the relation from \citet{IglesiasParamo2006}.  

\begin{table*}[]
\centering
\caption{Summary of Properties for UDGs with Tidal Features}
    \begin{tabular}{r|ll||lll}
    \hline\hline
         Property & KUG~0203-Dw1 & KDG~013 & UGC~9050-Dw1 & NGC~2708-Dw1 & NGC~5631-Dw1\\ \hline
         RA (J2000) & 02:05:46.5 & 02:08:33.6 & 14:09:13 & 08:56:12.7 & 14:26:13.6 \\
         Dec (J2000) & $-$09:50:30.8 & $-$07:50:59.2 & +51:13:28 & $-$03:25:14.8 & +56:31:50.2 \\
         $D$ (Mpc) & $26.8 \pm 1.9$ & $55.5 \pm 3.9$ & $35.2\pm 2.5$ & 40.6 & 28.4 \\
         $v_{\rm{helio}}$ (km/s) & $1885\pm1$ & $3856\pm2$ & $1952.1\pm0.3$ & \\
         $M_{V}$ & $-13.7\pm0.2$ & $-15.5\pm0.3$ & $-14.0\pm0.2$ & $-14.0\pm0.4$ & $-12.0\pm0.5$ \\
         $\log{\rm{L_{V}}/\rm{L}_{\odot}}$ & $7.4\pm0.1$ & $8.1\pm0.1$ & $7.5\pm0.1$ & $7.5\pm0.2$ & $6.7\pm0.2$ \\
         $(V-I)$ & $0.6\pm0.3$ & $0.8\pm0.1$ & $0.8\pm0.3$ & $0.8\pm0.3$ & $0.6\pm0.4$  \\
         $m_{g}$ & $18.5\pm0.1$ & $18.4\pm0.2$ & $18.9\pm0.2$ & $19.3\pm0.3$ & $20.5 \pm 0.4$ \\
         $(g-r)$ & $0.2\pm0.2$ & $0.4\pm0.3$ & $0.3\pm0.3$ & $0.5\pm0.4$ & $0.4\pm0.6$ \\
         $m_{NUV}$ & $19.1\pm0.4$ & $21.3\pm0.4$ & $19.9\pm0.2$ & -- & -- \\
         $m_{FUV}$ & $23.4\pm0.3$ & -- & -- & -- & -- \\
         $\mu(0,g)$ (mag arcsec$^{-2}$) & $25.5\pm0.3$ & $25.4\pm0.3$ & $\sim36.6\pm6$ & $24.9\pm0.6$ & $27.3\pm0.7$ \\
         $r_{1/2}$ (arcsec) & $11.8\pm1.8$ & $12.5\pm0.9$ & $\sim36.6\pm6$ & $13.2\pm2.9$ & $15.6\pm3.6$ \\
         $r_{1/2}$ (kpc) & $1.5\pm0.3$ & $3.4\pm0.3$ & $\sim6.2\pm1$ & $2.6\pm0.6$ & $2.2\pm0.5$ \\
         $b/a$ & $\sim0.5$ & $0.7\pm0.1$ & -- & $0.8\pm0.1$ & $0.5\pm0.1$ \\
         $\theta$ (deg) & 14 & 35 & -- & \\
         Sérsic index & $<0.52$\tablenotemark{a} & $0.6\pm0.1$ & -- & $1.2\pm0.2$ & 1.0\tablenotemark{b} \\
          $\log(M_{*}/M_{\odot})$ & $\sim7.1$ & $\sim8.0$ & $\sim$7.5 & $\sim7.5$ & $\sim6.4$ \\
         $\log(M_{\hi }/M_{\odot})$ & $\lesssim7.4$ & $\lesssim8.6$ & $8.4\pm0.0$ & $\lesssim7.3$ & $\lesssim7.2$ \\
         Projected distance (kpc) & 10.9 & 128.8 & $68.9 \pm 3.5$ & 45.2 & 34.1 \\
         SFR (NUV; \msun\ yr$^{-1}$ $\times10^{-3}$) & $6.2\pm2.2$ & $3.5\pm1.3$ & $5.1 \pm 1.0$ & -- & -- \\
         SFR (FUV; \msun\ yr$^{-1}$ $\times10^{-6}$) & $78\pm41$ & -- & -- & -- & -- \\
         Tidal features & stellar stream, & tidal plume & tidal tail & stellar stream & stellar stream \\
         ~ & \hi\ stream, & & \hi\ distortion & & \\
         ~ & tidal plume & & & & \\
        
    \hline
    \end{tabular}\\ [4pt]
    The left two UDG columns are the new data presented in this work. The right three UDG columns summarize the results reported in \citetalias{bennet2018}, \citetalias{jones2021}, and\citetalias{fielder2023} (specifically the ``Total'' column for UGC~9050-Dw1). Note that uncertainty estimates do not include the distance uncertainty. \\
    Rows: 1) RA: Right ascension in sexigesimal hh:mm:ss. 2) Dec: Declination in dd:mm:ss. 3) D: Assumed distance in Mpc 4) $v_{\rm{helio}}$: Heliocentric velocity. 5) $M_{V}$: $V$-band absolute Vega magnitude determined from CFHT data. 6) $\log{\rm{L_{V}}/\rm{L}_{\odot}}$: $V$-band luminosity. 7) $(V-I)$: $V-I$ color determined from CFHT data in Vega magnitudes. 8) $m_{g}$: $g$-band apparent AB magnitude, determined from CFHT data. 9) $(g-r)$: $g-r$ color determined from CFHT data, in AB magnitudes. 10) $m_{NUV}$: $NUV$ apparent magnitude determined from GALEX data, in AB magnitudes. 11) $m_{FUV}$: $FUV$ apparent magnitude determined from GALEX data, in AB magnitudes. 12) $\mu(0,g)$: $g$-band surface brightness in AB magnitudes. 13) \& 14) $r_{1/2}$: Circularized half-light radius (except for UGC~9050-Dw1, which is the radius at which the counts match the background). 15) $b/a$ Ratio of semi-minor to semi-major axis for the derived Galfit model. 16) $\theta$: Position angle of derived GALFIT model in degrees counterclockwise from north. 17) Sérsic index ($n$). 18) $\log({M_{*}/M_{\odot}})$: Logarithm of stellar mass using the M/L derived from \citet{zhang2017}. 19) $\log({M_{\hi\ }/M_{\odot}})$: Logarithm of \hi\ mass. 20) Projected distance from presumed host. 21) SFR: NUV derived star formation rate. 22) SFR: FUV derived star formation rate. \\
    \tablenotetext{a}{Because a double fit was required this value is based on the Sérsic fit of the core.}
    \tablenotetext{b}{The Sérsic index was fixed to 1 in this fit.}

    \label{tab:udg_props}
\end{table*}

\section{The Globular Cluster Systems}
\label{sec:gcs}

\subsection{Globular Cluster Selection}
\label{subsec:gc_selec}

GCs typically have radii of a few parsecs \citep{brodie2006}. Assuming a nominal radius of 5 pc, at the distance of KUG~0203-Dw1 and KDG~013 this corresponds to a radius of 0.04'' and 0.02'' respectively. The HST FWHM of the PSF is $\sim0.15''$ for ACS and $\lesssim 0.1''$ for WFC3, indicating that with either camera GCs will appear as point sources.

In this work, the globular cluster candidate (GCC) selection is identical to that described in \citetalias{jones2021} and \citetalias{fielder2023}. In brief, we perform aperture photometry on the HST images using the \textsc{Dolphot v2.0} software. We then perform subsequent cuts on the \textsc{Dolphot} produced point source catalog such that we limit to sources classified as `stars' (other source identifiers are `elongated object', `object too sharp', and `extended object'), with no photometric flags, and with S/N $> 5$. For both HST filters we further constrain with fit quality diagnostics. The sharpness is enforced to fall within the bounds of $-$0.3 to 0.3 and roundness is required to be less than 0.3, limiting elongated or distorted sources and those that are overly compact. We set a maximum limit on crowding to 0.5 mag in both filters. The \textsc{Dolphot} magnitude uncertainties are required to be less than 0.3 mag. Next a color cut is used, to limit sources to those within the color range expected for GCs. We use $0.5 < (V-I) < 1.5$ from \citet{brodie2006}, which should generally eliminate blue star-forming regions from our source selection.

Then we apply a concentration cut, which is determined by comparing the flux of the background subtracted $F814W$ image in concentric circular apertures of 4 and 8 pixels in diameter (see similar approaches in e.g., \citealt{peng2011,beasley2016,saifollahi2021}). We define our concentration index as $C_{4-8} = -2.5\log_{10}{(\frac{N_{4\rm{pix}}}{N_{8\rm{pix}}})}$ where $N_{4\rm{pix}}$ represents the sum of flux values in a 4 pixel aperture. To select GCCs we allow the concentration index parameter to span the range $0.2-0.8$ mag where we expect GCs to lie, which accounts for the extended locus of point sources (in magnitude-concentration space) and slightly extended sources. This is because larger GCs may not be perfect point sources, particularly at the distance of KUG~0203-Dw1. This selection is designed to eliminate diffraction spikes, compact background galaxies, and any other high concentration index sources.

Last, we employ a cut in magnitude in an effort to maximize the number of real GCs and eliminate contaminants. We do this by assuming the GCs follow the Gaussian dwarf elliptical GC luminosity function (GCLF) of \citet{miller2007}, similar to the approach of \citet{peng2011}. The Gaussian GCLF peaks at $mu_{M_{I},\rm{Vega}} = -8.12$ mag with $\sigma_{M_{I}} = 1.42$. We use the peak of the GCLF at the distance of each UDG to serve as a brightness \textit{minimum} in selecting our GCCs. This is used in lieu of the completeness limit in order to minimize stellar contaminants and sample parameter space where our data is effectively complete. We also employ a brightness maximum, above which we expect most point sources to be foreground stars. This value corresponds to $M_{I} > -12.38$, or $3\sigma$ from the mean of the GCLF. In apparent magnitude, we select all GCCs in the range of $19.8 < m_{I} < 24.0$ (where the 90\% completeness limit is 26) for KUG~0203-Dw1, and $21.3 < m_{I} < 25.6$ for KDG~013 (where the 90\% completeness limit is 26.3). Such a cut means we are only sampling half of the luminosity function, specifically the bright half of the GCLF. As a result, our final GC counts will be corrected by multiplying the initial resultant GC counts derived in this manner by 2. 

We provide a color-magnitude diagram in \autoref{fig:CMD} of point sources identified in \textsc{Dolphot}. All plotted points pass the initial set of signal-to-noise, sharpness, roundness, crowding, and magnitude uncertainty cuts, and red points pass the subsequent concentration cut. Point sources that fall within the color and magnitude selection criteria and lie within $2\times r_{1/2}$ constitute the final GCC sample. These sources are highlighted in the HST images in \autoref{fig:hst_img}, encircled by teal apertures. Note that in the KUG~0203-Dw1 frame (\autoref{fig:hst_img}, lower left) a majority of the GC candidates are centered on the two galaxies with few additional contaminants in the field, while the KDG~013 frame has substantially more contaminants (\autoref{fig:hst_img}, lower right).

\subsection{Globular Cluster Abundances}
\label{subsec:abundance}

After the \textsc{DOLPHOT} photometric catalog has been constrained following the procedure described above, we elect to select GCCs within $2\times$ the best fit circularized aperture radius, one of the most common choices within the literature for selecting GCs (most recently e.g., \citealt[][]{muller2021,danieli2022,janssens2022,jones2023}). While expanding to larger radii adds additional GC candidates, it also introduces a larger number of contaminants. In the case of KUG~0203-Dw1 $3\times r_{1/2}$ falls close to the host galaxy, meaning GCCs within this region could be associated with either galaxy. In the case of KDG~013,  $3\times r_{1/2}$ doubles the number of GCCs but this field has a rather high abundance of contaminants (see \autoref{fig:hst_img} lower right). We therefore take the standard approach of using $2\times r_{1/2}$ to mitigate contaminants. 
In \autoref{subsec:optical_props} note that the best GALFIT models had some ellipticity. Selecting GCCs within an elliptical aperture yielded similar counts to those selected within a circularized aperture. Likewise, due to the tidal features of these UDGs,  defining a strict boundary for these galaxies is non-trivial. We therefore choose to adopt the simpler approach of counting GCCs within a general circularized radius and subsequently subtracting contaminants.

Contaminant counts for false GCCs were determined with a parallel WFC3 field taken in an empty patch of sky close to the host systems. This is particularly useful for KUG~0203-Dw1, where the host lies in the frame making it difficult to disentangle GCs associated with the host and actual contaminants. Additionally, the coordinates of the parallels are checked to ensure they are sampling similar environments to that of the UDG. To determine contaminant rates we employ the same cuts used to identify GCCs in the ACS field as described in \autoref{subsec:gc_selec}. The WFC3 field-of-view (160''x160'') is smaller than the ACS field-of-view (202''x202''), so the number of contaminants identified must be scaled relative to the ACS detector sky area incorporating the pixel size difference before scaling to within the area of the actual GC searches. We note that contaminant counts were also computed for the ACS pointings, which yield comparable GC abundances.

The raw GCC counts within $2\times$ the effective radii of either UDG are small ($<10$ per UDG), where number counts are dominated by Poisson statistics. We thus rely on a Bayesian approach to determine the number of GCs within the UDG aperture analytically. The procedure for this method is outlined in Section 4.3 of \citetalias{jones2021}, which we summarize here. The count of potential false GC candidates within the UDG's aperture conforms to a Poisson distribution with an unknown mean, denoted as `\textit{b},' representing the background rate of false positives arising from GCs linked to the host galaxy and other incidental sources such as Milky Way foreground and compact background galaxies. The probability of false GCCs within the aperture of the UDG can be described by a normal Poisson process. Multiplying this probability with the analytic Bayesian solution with a flat prior to estimate $b$ and $b$ marginalized yields the probability mass function for the number of GCs. This result is written as: 

\begin{equation}
    p(N_{\rm{GC}}|O) = \frac{\Gamma(N_{\rm{bg}}+N_{\rm{GCC}}-N_{\rm{GC}}+1)}{(A_{\rm{bg}}+1)^{(N_{\rm{bg}}+N_{\rm{GCC}}-N_{\rm{GC}}+1)}(N_{\rm{GCC}}+N_{\rm{GC}})}, 
\label{eq:prob}
\end{equation}
where $O$ is short for observations, $N_{\rm{GC}}$ is the number of genuine GCs in the UDG apertures, $N_{\rm{GCC}}$ is the number of GCCs in the UDG aperture, $N_{\rm{bg}}$ is the number of background sources that pass as false GCCs, and $A_{\rm{bg}}$ is the area of the background field. $A_{\rm{bg}}$ is dimensionless as it is expressed in terms of the area of the circular aperture used to count GCCs around the target. Because we use the WFC3 field to determine background counts, $A_{\rm{bg}}$ is scaled such that it is in units of the ACS area. 

While the WFC3 parallels are close to the systems of interest, we may still suffer from small number statistics. Thus we determine an estimate of contaminants using the web-based tool  \textsc{TRILEGAL v1.6} \footnote{\url{http://stev.oapd.inaf.it/cgi-bin/trilegal_1.6}} \citep{girardi2005,girardi2012}. \textsc{TRILEGAL} (TRIdimensional modeL of thE GALaxy), a model of Milky Way star counts, constructs a three-dimensional representation of our galaxy, incorporating various stellar populations in its thin disk, thick disk, halo, and bulge components calibrated using extensive survey data. Employing TRILEGAL, we generate a simulated star catalog with default settings and a Chabrier log-normal initial mass function \citep{chabrier2001}. We query a full square degree centered on each galaxy, and then randomly down-sample 10,000 times to both the area of the ACS field and the WFC3 field to ensure our contamination estimates are consistent. We find similar estimates to those obtained from the WFC3 fields.

The search radius of KUG~0203-Dw1 is equivalent to $23.6\pm3.6$ arcsec or $3.1\pm0.6$ kpc at its distance (see \autoref{tab:udg_props}). A total of 5 GCCs were identified within this aperture (see the outer dashed radiis in \autoref{fig:hst_img}). We find a total contaminant abundance in the parallel field of 13, which when scaled to the ACS area covered by the UDG yields a raw abundance of 1 contaminant. For comparison \textsc{TRILEGAL} estimates $3\pm2$ contaminants. With our probabilistic approach for determining GC abundance and multiplying our result by $2$ to account for only sampling the bright half of the GCLF, our final most probable GC abundance for KUG~0203-Dw1 is $N_{\rm{GC}} = 8\pm2$. 

The search radius of KDG~013 is $25.0\pm$1.8 arcsec or $6.7\pm0.5$ kpc . Within this aperture we identify a total of 4 GCCs, with a raw contaminant abundance of 2 in the UDG area. \textsc{TRILEGAL} estimates $2\pm1$ contaminants. After performing our probabilistic analysis we find our most probable GC abundance for KDG~013 is $N_{\rm{GC}} = 6\pm2$. 

We can extend the $I$-band magnitude cut in \autoref{subsec:gc_selec} from $M_{I} < -8.12$ down to our $90\%$ completeness limit, which corresponds to $M_{I} < -6.14$ for KUG~0203-Dw1 (sampling about 2/3 of the faint-end GCLF) and $M_{I} < -7.42$ for KDG~013 (sampling nearly the full GCLF). While a looser cut such as this may allow for an increase in contaminants, we find the same predicted abundance for both systems (within error) that we do when determining the GC abundance from the bright half of the GCLF alone. At this extended magnitude limit we find $9^{+1}_{-2}$ GCs for KUG~0203-Dw1 and $8\pm2$ for KDG~013.

\begin{figure*}
    \centering
    \includegraphics[width=0.8\linewidth]{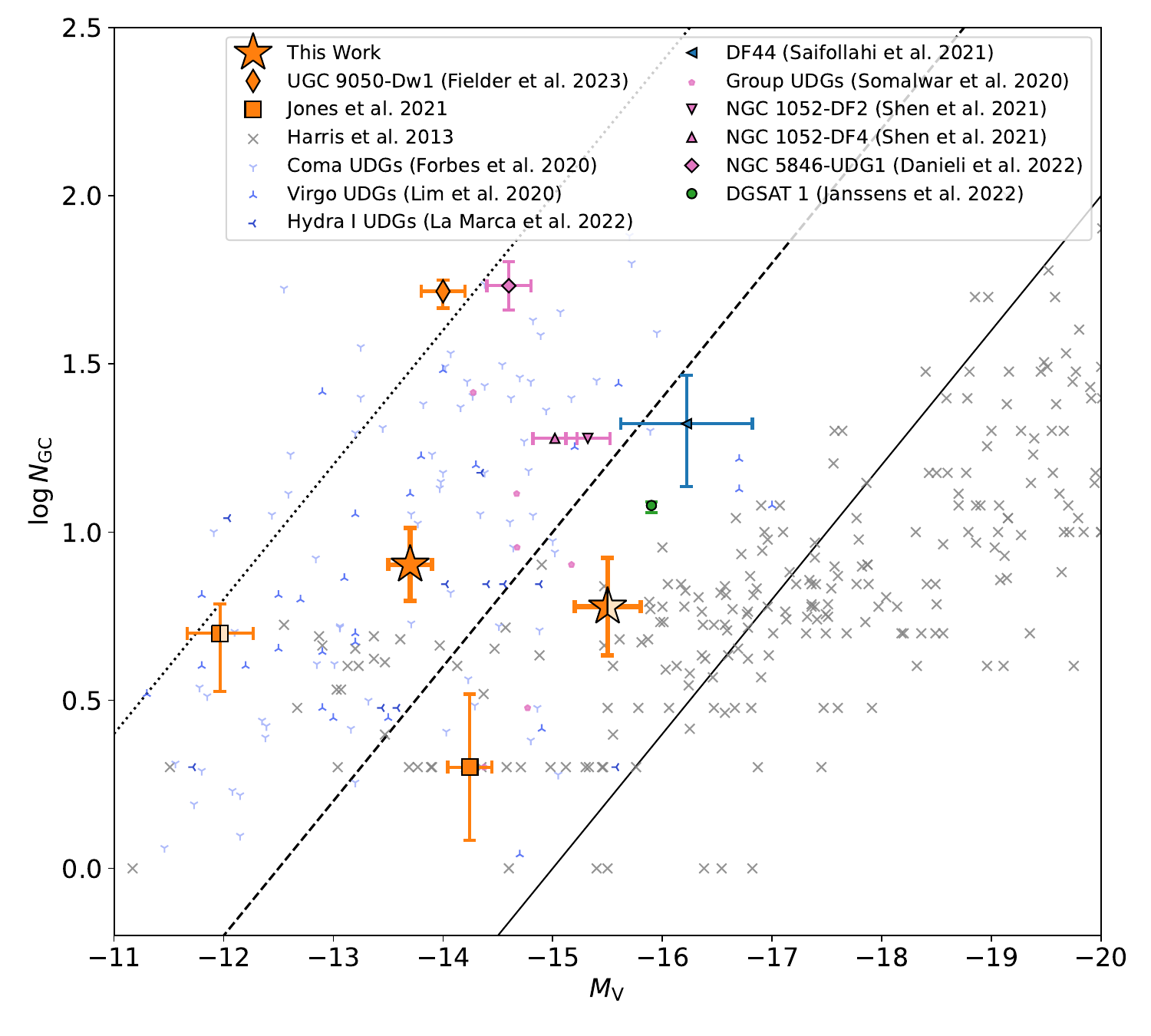}
    \caption{Counts of GCs ($N_{\rm{GC}}$) plotted against the $V$-band absolute magnitude ($M_{V}$). To provide context, we compare our findings to several data sets found in the literature, including a general galaxy data set (grey x's from \citealt{harris2013}), cluster UDGs (various blue points), group UDGs (various pink points), and other distinct GC-rich UDGs from various environments. The solid, dashed, and dotted black lines represent specific frequency ($S_{N}$) values of 1, 10, and 100, respectively. Our full sample of five UDGs with tidal features is shown by orange points: KUG~0203-Dw1 (star), KDG~013 (half filled star), UGC~9050-Dw1 (diamond), NGC~2708-Dw1 (square), and NGC~5631-Dw1 (half filled square). KUG~0203-Dw1 and KDG~013 are slightly more GC rich than the two UV-bereft UDGs in our sample (NGC~2708-Dw1 and NGC~5631-Dw1; \citetalias{jones2021}).}
    \label{fig:NGC_MV}
\end{figure*}

The GC abundances for these systems are plotted in \autoref{fig:NGC_MV} (orange stars) alongside the rest of our sample of UDGs with tidal features (other orange points). The nearby galaxy sample of \citet{harris2013} is plotted for comparison, which includes elliptical, spiral, lenticular, and irregular galaxies and extends to brighter galaxies. We also include a number of UDG samples from both clusters (blue points) and groups (pink points) for comparison: Coma cluster UDGs \citep{forbes2020}, Virgo cluster UDGs \citep{lim2020}, Hydra I cluster UDGs \citep{iodice2020,lamarca2022}, and group UDGs compiled in \citet{somalwar2020}. There are also a number of individual UDGs studied in the literature that we highlight, which are considered to be GC-rich or contain over-luminous GCs (DF2, DF4, DGSAT~1): Dragonfly 2 (DF2) and Dragonfly 4 (DF4) (NGC~1052 UDGs; \citealt{shen2021}), Dragonfly 44 (DF44, a Coma UDG; \citealt{saifollahi2021} updated from the \citealt{forbes2020} values), 
NGC 5846-UDG1/MATLAS19 (NGC~5846 group UDG; \citealt{muller2021,danieli2022}), DGSAT~1 (in a low density region of the Pisces-Perseus supercluster; \citealt{janssens2022}). 

A study of local volume dwarfs performed by \citet{carlsten2022} finds that in the stellar mass range of $7 < \log{M_{*}/M_{\odot}} < 8$ (the approximate mass range of these UDGs) the typical GC abundance ranges from approximately $1-7$. Specifically, the Fornax dwarf ($\log{M_{*}/M_{\odot}} \sim 8$) hosts $6$ bright GCs \citep{deBoer2016,pace2021} at $\sim1.5$ magnitudes below the peak of the GCLF. These results for local dwarfs are very comparable to the values we derive for KUG~0203-Dw1 and KDG~013, although both UDGs are close to the expected upper limit.

\subsection{Specific Frequency}
\label{subsec:sf}

A convenient way to describe GC richness relative to the galaxy's luminosity is to calculate specific frequency. While this quantity is common to calculate even at dwarf galaxy masses, the relation between GC abundance and luminosity does not hold as well as it does at higher masses. This is speculated to be due to the stellar-to-halo mass relation becoming less linear at low mass, as GCs are found to trace total DM halo mass and luminosity traces stellar mass (hence why at higher masses a specific frequency of 1 is more common). The definition of specific frequency presented in \citet{harris1981} is $S_{N} = N_{\rm{GC}}10^{0.4(M_{V}+15)}$. Using our derived galaxy $M_{V}$ from the CFHT data and the GC abundances this yields a specific frequency of $S_{N} = 26.9_{-8.8}^{+8.2}$ for KUG~0203-Dw1 and $S_{N} = 3.3_{-0.6}^{+0.8}$ for KDG~013. 

In \autoref{fig:NGC_MV} we include diagonal lines of constant specific frequency. KUG~0203-Dw1 and KDG~013 both have specific frequencies consistent with other UDGs and dwarf galaxies (generally between 1 and 50).

\section{The Formation of KUG~0203-Dw1 and KDG~013}
\label{sec:formation}

In this section we comment on the possible formation scenarios of KUG~0203-Dw1 and KDG~013. As both UDGs exhibit clear tidal features we will be considering formation mechanisms that are likely to produce such features or could occur to dwarf galaxies in group environments, such as: 1) Dwarf mergers, 2) TDGs, 3) Dwarf galaxies ``puffed up'' by interactions with a more massive galaxy, and 4) pre-existing UDGs subsequently processed by the group environment. Our approach is to find the simplest formation method that is the most consistent with the observed properties of either UDG. Thus we favor the method where we must invoke the minimal number of processes to explain the diffuse nature of these galaxies. However, it is impossible to completely rule out any formation scenario.


\subsection{Dwarf-Dwarf Interactions}

When dwarf galaxies undergo collisions, the result can manifest as an increase in the effective radius and angular momentum of the system, accompanied by a decrease in surface brightness and a redistribution of star formation towards the outer regions of the resulting (merged) galaxy. 
Dwarf major mergers give rise to distinctive tidal features such as tails, bridges, and shells, expected to last $2-5$~Gyrs \citep[e.g.,][]{johnston2001,pearson2018}. But distinguishing effects caused by tidal interactions with a massive companion from those caused by a dwarf merger is often challenging, aside from when the dwarfs are not completely merged \citep{paudel2018}.

There are observed properties in these UDGs that directly challenge the dwarf merger theory. A hydrodynamical simulation study (ROMULUS25) conducted by \citet{wright2021} suggests that decentralized star formation leads to a steep color gradient. However, the UV emission in KUG~0203-Dw1 and KDG~013 is evenly distributed throughout either galaxy meaning there is no de-centralized tracer of star formation, and we detect no notable color gradient in the optical.  

Additionally, the \hi\ morphology observed in KUG~0203-Dw1 is a direct result of the interaction with KUG~0203-100 (see \autoref{subsec:hi_props}). While it is possible a dwarf merger occurred before the system began interacting with the host, we choose not to invoke multiple formation mechanisms for this UDG to avoid over-complication. It is plausible that KDG~013 resulted from a more recent gas-poor merger. However, given that no current simulations exist for comparison to recent (z $<1$) dwarf mergers yielding UDGs, as nearly all UDGs formed via mergers in the ROMULUS25 simulations occur at earlier times, a recent merger is difficult to assess. While bullet cluster-like collisions have also been suggested to produce UDGs \citep[e.e.g,][]{ogiya2022, vandokkum2022}, this formation mechanism is still not well understood and the observed properties of our UDGs are not consistent with those suggested to have formed from such a mechanism, such as an over-abundance of GCs. Therefore we suggest that the dwarf major merger formation mechanisms is unlikely for either UDG.

\subsection{Tidal Dwarf Galaxies}

Tidal dwarf galaxies (TDGs) form as a result of tidal interactions between two or more massive galaxies, where the tidal forces generated during close encounters lead to the ejection of stellar and gaseous material into extended tidal tails or bridges. These ejected materials can then coalesce and form new, self-gravitating galaxies known as TDGs \citep[e.g.,][]{elmegreen1993,duc1998,kaviraj2012}. Note that TDGs and UDGs with tidal features (sometimes called tidal UDGs) refer to different classes of objects, which in some instances may or may not overlap. 

\citet{bournaud2006} find in their TDG formation simulations that tidal streams responsible for forming a TDG typically dissolves within $300-500$~Myr. Additionally, the \hi\ streams are expected to persist longer than stellar streams \citep{jones2019,lee-waddell2019,namumba2021}. KUG~0203-Dw1 has been shown to have a clear \hi\ stream while KDG~013 does not, meaning that we might expect if they are TDGs then KUG~0203-Dw1 is $<500$~Myr old and KDG~013 is $>500$~Myr old.
While long-lived TDGs are anticipated to form at the tips of \hi\ tails \citep{bournaud2006}, KUG~0203-Dw1's central location within the stream, with leading and trailing tails more indicative of dwarf stripping, suggests otherwise. 

Because TDGs form very differently from typical galaxies, the expectation is that TDGs do not contain GCs (or DM), and at present, studies of low redshift TDGs do not find GCs \citep[e.g.,][]{fensch2019}. From our Bayesian analysis discussed in \autoref{subsec:abundance} we can obtain probabilities that either UDG has 0 real GCs. In the case of KUG~0203-Dw1 there is only a $0.75\%$ probability that none of the GC candidates are real, with a $7.26\%$ probability for KDG~013. Hence, the abundances of GCs provide strong evidence against categorizing these UDGs as young TDGs.

It has been suggested that long lived GCs could form in an old TDG \citep{fensch2019}, but current observational evidence is lacking on this front. Therefore we posit that given our current understanding of TDG formation and the supporting observational evidence, a TDG origin is unlikely for either UDG, especially given their GC abundance.

\subsection{Tidally `Puffed Up' Dwarfs}
Strong gravitational effects induced by tidal interactions with massive companions can expand a galaxy's stellar halo, thereby decreasing the central surface brightness and pushing a dwarf galaxy into UDG parameter space  \citep{errani2015,conselice2018,carleton2019,tremmel2020}. While previous research has primarily focused on UDGs in cluster environments \citep[e.g.,][and sources mentioned above]{sales2020,carleton2021}, there is mounting evidence indicating that this phenomenon may also occur in lower density group environments.
Hydrodynamical simulations by \citet{jiang2019} and \citet{liao2019} suggest that approximately 50\% of group UDGs form from dwarfs on eccentric orbits undergoing tidal stripping and heating, with the remainder being pre-existing UDGs later affected by group dynamics. After a pericentric passage the UDG progenitor is stripped of a substantial portion of its gas and star formation slows \citep{roman2017, benavides2021}. \citet{jiang2019} and \citet{liao2019} find that within 1 to 2 pericentric passages a dwarf is transformed into a UDG and has lost its cold gas. 

Both KUG~0203-Dw1 and KDG~013 have tidal plumes, disturbed morphologies, and KUG~0203-Dw1 has a clear stellar and \hi\ stream connection to its host galaxy. It is without doubt that the group environment must have played a role in shaping the way the UDGs appear now. Additionally, we have demonstrated both KUG~0203-Dw1 and KDG~013 display GC abundances typical of galaxies given their $V$-band brightness (\autoref{fig:NGC_MV}), suggestive of a classical dwarf origin.

KDG~013 has no clear \hi\ detection, implying that NGC~830 or other group members may have pulled its gas off after pericentric passage, instigating its UDG nature. The situation is less clear for KUG~0203-Dw1, which still contains an over-density of \hi\ and is in close proximity in projection to the host (only 10.9~kpc). Thus the UDG interaction with the host is likely in an early phase, where the UDG is being tidally puffed but has not yet been fully stripped of its gas. This is further motivated by the rather small circularized radius of the UDG, which just barely passes the UDG criteria ($1.53\pm0.3$~kpc compared to the limit of $>1.5$~kpc). Because KUG~0203-Dw1 has a bluer stellar population and high specific star formation rate than KDG~013 one can envision them at different stages in the tidal processing in the group, with KUG~0203-Dw1 just forming into a UDG and KDG~013 already having been processed. 

This is the formation mechanism that favors our GC abundance most, and that explains both the observed tidal features and the UDG nature of the galaxies without having to invoke additional UDG formation mechanisms in addition to tidal processing within the galaxy group.

\subsection{Pre-existing UDGs Processed By the Group}

While half of the group UDGs in the simulations of \citet{jiang2019} and \citet{liao2019} originate as dwarf galaxies where subsequent tidal effects transformed them into UDGs, the other half originate as pre-existing UDGs in the field or group periphery that then experience environmental tidal effects as they fall in. This would mean the tidal features are coincidental and unrelated to the diffuse nature of the UDGs. 

Differentiating between a pre-existing UDG and a UDG resulting from group processing is non-trivial in these simulations. While we cannot completely rule out that KUG~0203-Dw1 and KDG~013 were pre-existing UDGs, our observations provide hints that this is less likely. The observation that KUG~0203-Dw1 narrowly meets the UDG criteria suggests that it might not have originally qualified as a UDG before interacting with its host. Likewise, we may expect pre-existing UDGs to become much more extended after interacting with a group, but KUG~0203-Dw1 and KDG~013 are both approximately average in size for UDGs in terms of half light radius (refer to \autoref{fig:size_lum} in the following section). For example the pre-existing UDG explored in \citet{liao2019} has a half light radius more than double that of the UDG formed by tidal heating.



Currently the lack of specific predictions for GC abundances in hydrodynamical simulations complicates distinguishing between UDG formation via tidal heating and pre-existing UDGs that suffered tidal heating. \citet{kong2022} compared gas-rich field UDGs with hydrodynamical simulations, finding discrepancies between simulated and observed features. This implies that current galaxy formation models may lack accuracy for gas-rich UDGs. In addition, the field UDGs later accreted in the simulations by \citet{jiang2019} and \citet{liao2019} differ in their formation mechanisms (star formation feedback versus high spin halos). Thus a more comprehensive understanding of group UDGs is warranted before conclusive arguments for the processing of pre-existing UDGs can be made. 

\smallskip

We contend that the stellar populations in both KUG~0203-Dw1 and KDG~013 are distinctly influenced by their group environments. 
We favor the tidally ``puffed-up'' origin as it explains both the tidal features and the UDG nature of these galaxies without having to employ a separate UDG formation mechanism in addition to group processing. However, presently, origins as pre-existing UDGs cannot be conclusively dismissed. Perhaps velocity measurements, akin to those conducted for UDGs in the Coma cluster by \citet{alabi2018}, are necessary for such distinction, especially if current hydrodynamical simulations prove accurate.

\section{Comparative Analysis of the UDG Sample}
\label{sec:comparison}

\subsection{The Sample}
\label{subsec:sample}

In addition to the two UDGs presented here, we also have previously reported an additional 3 UDGs that were identified and observed in an identical manner, constituting a full sample of 5 UDGs. These UDGs were selected such that they are a) in group-like environments within proximity to a more massive `host' galaxy and b) have telltale signs of tidal disturbances through elongated or amorphous morphology, tidal tails or plumes, and/or stellar streams. These are all of the UDGs in the wide portion of the CFHT Legacy survey covering 150 deg$^{2}$ that have been identified to contain clear evidence of interaction given their morphology and/or the presence of stellar streams. Of this set of five, two of the UDGs - NGC~2708-Dw1 and NGC~5631-Dw1 - contain little evidence of recent star formation or neutral gas and exhibit slightly redder optical colors. The other three - UGC~9050-Dw1, KUG~0203-Dw1, and KDG~013 - in contrast have GALEX UV detections which is suggestive of recent star formation ($<100$~Myr) and also exhibit bluer optical colors (although the optical color errors are large). We briefly summarize the previously reported results of the first three UDGs within our sample below before comparing the set of five. The properties of these UDGs are also presented in \autoref{tab:udg_props}. This is the first consistently identified and measured sample of UDGs exhibiting tidal features.

\subsection{Targets Previously Reported}

NGC~2708-Dw1 is an elongated UDG companion to NGC~2708 (of the NGC~2698 group), with a faint linear stellar stream connecting the two.
\citetalias{jones2021} identified $N_{\rm{GC}} = 3^{+1}_{-2}$ and $S_{N}=3.3^{+5.6}_{-1.0}$ for NGC~2708-Dw1. Additionally there is no trace of \hi\ in the UDG itself or of a \hi\ tidal connection to the host NGC~2708, indicating that the stellar stream originated from a gas poor object (likely the UDG itself) and not the disk of NGC~2708.
 
NGC~5631-Dw1 is more notably elongated, connected to NGC~5631 with a faint curved stellar stream. \citetalias{jones2021} finds a slightly higher GC abundance, with $N_{\rm{GC}} = 5^{+1}_{-2}$ and a much higher specific frequency of $S_{N}=54^{+49}_{-33}$. There is a plausible \hi\ detection coincident with the UDG and connected to the disturbed \hi\ content of NGC~5631. However, because the \hi\ tails are significantly larger than the stellar tails, they may not be physically connected nor associated with the UDG. 

UGC~9050-Dw1 is a highly amorphous UDG with a distinct tidal tail, in the proximity of UGC~9050, but with no clear stellar or \hi\ streams between the two. It contains a brighter ``core'' region with overlapping UV emission while the remainder of the galaxy consists of very low surface brightness and redder stellar material. \citetalias{fielder2023} reports an extraordinary GC population of $N_{\rm{GC}} = 52\pm12$ and $S_{N}=122\pm38$. 
The colors of the GCs exhibit a narrow spread, indicating that the vast majority of them likely formed in a narrow window in time. This UDG has a clear \hi\ detection, with an estimated mass of $\log{M_{\hi\ }/M_{\odot}} = 8.44\pm0.04$.

\subsection{Comparisons Between UDGs With Tidal Features}
\label{subsec:full_sample_comparison}

\begin{figure*}
    \centering
    \includegraphics[width=\linewidth]{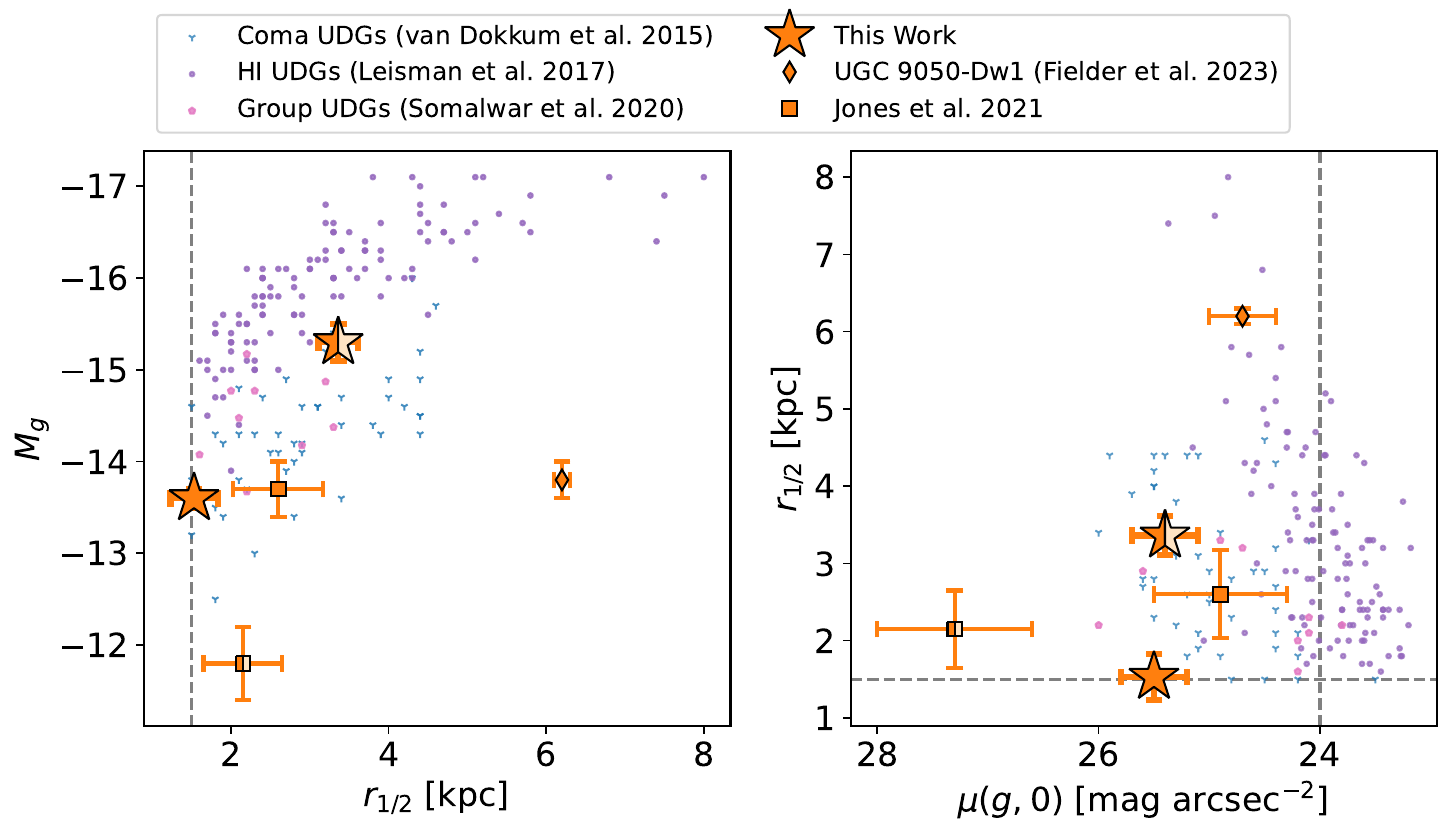}
    \caption{\textit{Left:} Size-luminosity relation for the sample of 5 UDGs with tidal features compared to Coma cluster UDGs (\citealt{vandokkum2015}; blue triquetra), group UDGs (\citealt{somalwar2020}; pink pentagons), and \hi\--rich UDGs (\citealt{leisman2017}; purple points). All of our UDGs are consistent with the Coma sample, aside from UGC~9050-Dw1 which is unusually large for its brightness. The dashed vertical line marks the defined UDG lower limit on size used in this work.
    \textit{Right:} Central surface brightness as a function of half light radius for the same sample. Here UGC~9050-Dw1 is more consistent with the \hi\ sample while NGC~5631-Dw1 has a very faint central surface brightness. We note that the comparative samples are surface brightness limited so there may be other sources currently undetected at lower central surface brightness. The dashed horizontal line is analogous to that in the left panel. The dashed vertical line indicates the maximum central surface brightness for a UDG as defined in this work (please note that the \hi\ UDGs adhere to a distinct definition hence why some are brighter).
    }
    \label{fig:size_lum}
\end{figure*}

\textbf{Physical Properties} 
The five UDGs roughly span the physical parameter space with diverse luminosities and stellar/\hi\ masses. We illustrate the size, luminosity and central surface brightness in \autoref{fig:size_lum}, comparing to UDG samples from the Coma cluster \citep{vandokkum2015}, group environments \citep{somalwar2020}, and \hi-bearing UDGs \citep{leisman2017}. UGC~9050-Dw1 is a clear outlier in the size-luminosity relation ($\sim6$kpc in radius, though this is not the half-light radius because of the complex morphology of the UDG), with the remaining four UDGs consistent with the Coma cluster and group UDGs. NGC~2708-Dw1, NGC~5631-Dw1, KUG~0203-Dw1, and KDG~013 have comparable dimensions, with half-light radii ranging from 1.5 to 3.4~kpc, although KUG~0203-Dw1 just meets the UDG criteria. In terms of central surface brightness-half light radius parameter space, UGC~9050-Dw1 somewhat resembles the \hi\ UDG population (although still rather large) while NGC~5631-Dw1 has a very faint central surface brightness. However, previous studies limited by surface brightness may not have thoroughly explored this parameter space, rather than NGC~5631-Dw1 being a significant outlier. We also note that these other samples tend to assume $n=1$ for the Sérsic index, which complicates this comparison.

\textbf{Optical Morphology} Our sample exhibits a variety in its optical morphology, not atypical for dwarf galaxies. NGC~2708-Dw1, NGC~5631-Dw1, and KUG~0203-Dw1 have associated stellar streams connecting to a massive host. UGC~9050-Dw1 has a very clear tail feature extending to the north of the UDG with more subtle material extending southward, and KUG~0203-Dw1 and KDG~013 each have a tidal plume extended towards the southwest.

\textbf{Stellar Populations} 
While we cannot make detailed assessments of the stellar populations within the UDGs, we can comment on the ages of the stellar population given the presence of UV emission. The UV emission from short-lived, high-luminosity stars, with lifetimes $<100$ Myr, peaks at ultraviolet wavelengths, highlighting recent star formation activity \citep{salim2014, tuttle2020}. KUG~0203-Dw1, UGC~9050-Dw1, and KDG~013 all have UV emission detected by GALEX (in order of SFR). Thus we expect that these UDGs have younger stars than NGC~5631-Dw1 or NGC~2708-Dw1. Notably only the very central region of UGC~9050-Dw1 contains UV emission, meaning that the center of the galaxy may have a younger stellar population that the outskirts.   

\textbf{Gas Content}
Only two of the UDGs have clear neutral gas detections: KUG~0203-Dw1 ($\sim10^{7}~\msun$) and UGC~9050-Dw1 ($2.75\times10^{8}~\msun$). There are detections coincident with NGC~5631-Dw1 and KDG~013 but because there is no statistically significant concentrations of \hi\ at the location of the galaxies, it is not clear that the \hi\ is associated with them. 

The morphology of the \hi\ of KUG~0203-Dw1 shows both a connection between the UDG and the host, in addition to a spur to the southeast of the UDG and a trailing/leading tail detection to the northeast of the host. This suggests that the system initially contained gas, which is currently undergoing tidal stripping by the host galaxy, rather than the UDG stripping gas from the host. The latter would not lead to a pronounced concentration of \hi\ centered on the UDG. In contrast, while the \hi\ in UGC~9050-Dw1 is also distorted, it is not apparent if this is due to its neighbor UGC~9050. UGC~9050 does have an \hi\ spur extended in the direction of UGC~9050-Dw1, possibly indicative of a prior interaction.

\textbf{GC Populations} 
In comparison to the two UV bereft UDGs included in our sample (NGC~2708-Dw1 and NGC~5631-Dw1 from \citetalias{jones2021}), our sample of UV bright UDGs KUG~0203-Dw1 and KDG~013 contain similar but slightly more abundant GC populations (see \autoref{fig:NGC_MV}). All four have abundances comparable to normal galaxies at their luminosity (see \autoref{subsec:abundance}), although KUG~0203-Dw1 and KDG~013 are at the upper end. In general, cluster UDGs lie above group UDGs in parameter space (not unexpected based on theoretical work, e.g., \citealt{carleton2021}) with more elevated GC abundances. However, group UDGs seem to display a rather wide variety in GC abundances overall, especially given that of UGC~9050-Dw1 and the more GC abundance UDGs identified in \citet{somalwar2020} and the NGC~1052 group \citep{shen2021,vandokkum2022}.

\begin{figure*}
    \centering
    \includegraphics[width=\linewidth]{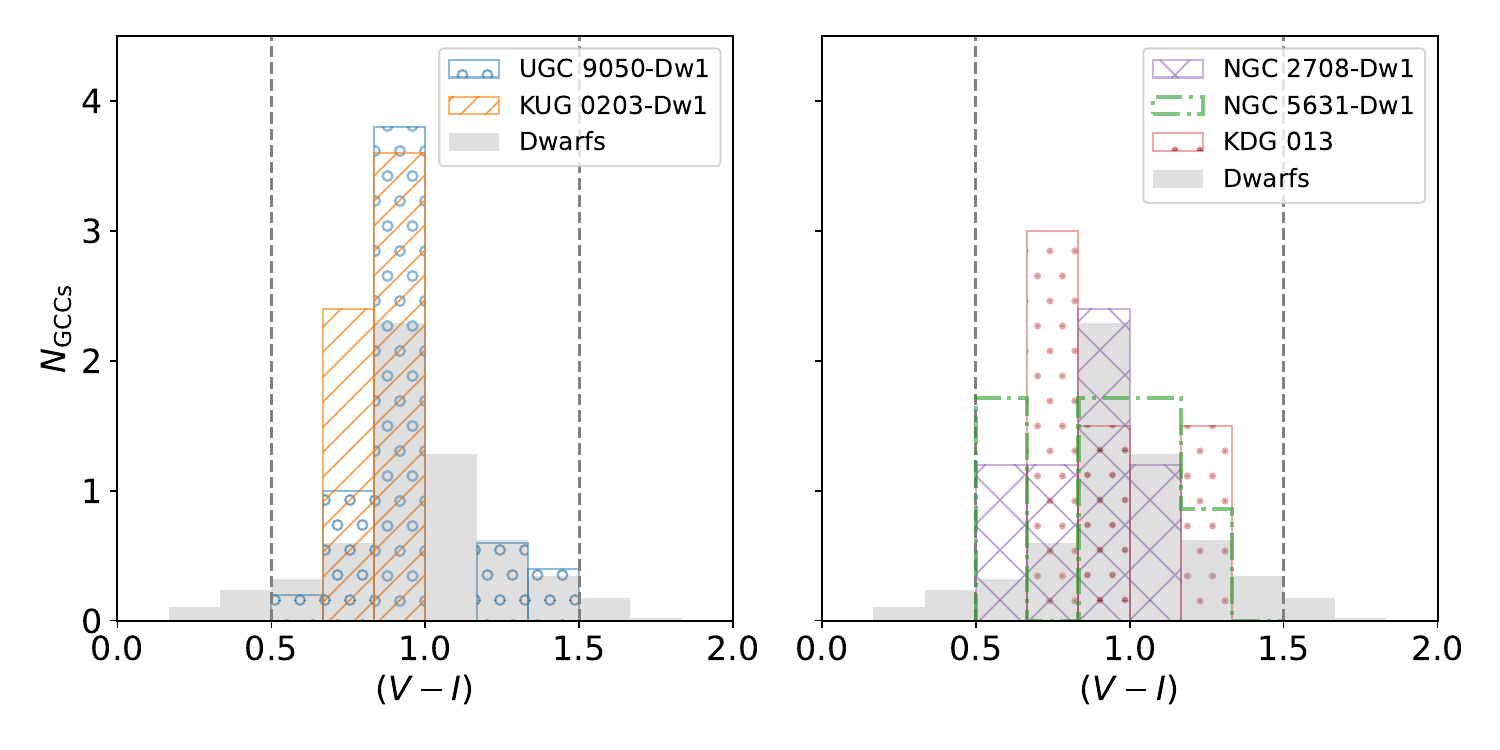}
    \caption{$(V-I)$ color distribution of the 5 UDGs with associated tidal features. The sample is split for readability. The left histograms shows the UDGs with GCs exhibiting a narrow spread in color and the right histogram shows the other 3 UDGs. We also include samples of dwarf galaxies  for comparison \citep[from ][]{sharina2005,georgiev2009}. Each histogram is normalized such that it has an integrated area of 1 such that the histograms can be compared on similar axes (hence non integer GC abundances). The vertical dashed lines correspond to the color cut employed for GC selection; $0.5<(V-I)<1.5$. KUG~0203-Dw1 exhibits a blue GC population, similar to UGC~9050-Dw1. The GC candidates of NGC~2708-Dw1, NGC~5631-Dw, and KDG~013 generally span the full color space. }
    \label{fig:GC_color}
\end{figure*}

We plot the colors of the GCs of our sample in \autoref{fig:GC_color}. Note that the histograms have been normalized to have an integrated area of 1 due to the high GC abundance of UGC~9050-Dw1. Three of the UDGs in our sample have GC colors exhibiting a broad spread within the allowed $(V-I)$ color, as expected for the general UDG and dwarf galaxy population \citep[see e.g.,][]{sharina2005,georgiev2009}. Results in \citetalias{fielder2023} show that a vast majority of the GCs identified in UGC~9050-Dw1 exhibit a narrow color spread ($\sigma_{(V-I)}$ $0.08$ mag) and are blue, with a very small redder GC candidate population consistent with contaminants. KUG~0203-Dw1 also appears to have a GC population with a relatively uniform and blue color, with mean $(V-I) = 0.83$ mag, median $(V-I) = 0.85$ mag, and $\sigma_{(V-I)}$ $0.08$ mag. Given that the field KUG~0203-Dw1 resides in has a small number of contaminants it is not likely that any of the GC candidates are foreground stars or background galaxies. However, because the GC abundance is small in KUG~0203-Dw1, this could also be a chance effect and we hesitate to draw significant conclusions based on the GC colors here. In the literature the distinction between UDGs with uniform and non-uniform GC colors persists regardless of tidal influences (possibly tidal: NGC~1052-DF2, NGC~1052-DF4 see \citealt{vandokkum2022}; not tidal: NGC~5846-UDG1, DGSAT~1 see \citealt{muller2021,janssens2022} respectively). 

\textbf{Environment} All of the UDGs studied in this work reside in group environments. Each of the groups have a confirmed membership of $n\gtrsim 4$ except for UGC~9050-Dw1, which currently has only two known members (note that the UGC~9050 pair lies at the fringe of a larger group, although association is unclear). This is also the only group where the host galaxy has a similar mass (only one order of magnitude different in stellar mass and half an order of magnitude different in \hi) to the UDG. This environment may be connected to the more unusual properties exhibited by the UDG. KUG~0203-Dw1 and KDG~013, though smaller, exhibit sizes comparable to the other members of their respective systems. In contrast, NGC~2708-Dw1 and NGC~5631-Dw1 are significantly dwarfed by their companions. Possibly, the stronger influence of the host on these smaller systems led to an earlier period of gas stripping. This, in turn, resulted in a smaller GC abundance and a lack of UV emission. 

\textbf{Plausible Formation mechanisms} \citetalias{jones2021} concludes that  NGC~2708-Dw1 and NGC~5631-Dw1 are most consistent with a tidally ``puffed up'' dwarf formation origin, given that they have GC abundances comparable to typical dwarf galaxies and both have stellar streams connecting to their hosts. We draw similar conclusions for the UDGs studied in this paper. While we cannot conclusively dismiss that any of these UDGs were UDGs prior to interacting with their group environments, the GC abundances, half-light radii, and other properties favor a group processing origin. Additionally, this formation scenario explains both the UDG nature of these objects and the tidal features, without invoking multiple formation mechanisms simultaneously. In contrast, \citetalias{fielder2023} concludes that UGC~9050-Dw1 is most consistent with a dwarf major merger formation mechanism given its large size, abundant and relatively uniform in color GC population, and other unusual characteristics. These varying findings demonstrate that multiple formation mechanisms for UDGs in group environments are viable, highlighting that UDGs form in different ways.



\section{Conclusion}
\label{sec:conclusion}

We present HST imaging and VLA \hi\ mapping to follow-up two group UDGs with elongated morphologies and tidal features, indicative of a tidal disturbance. We find the GC counts of both KUG~0203-Dw1 ($8\pm2$) and KDG~013 ($6\pm2$) to be consistent with normal dwarf galaxies of similar luminosities (though on the upper end). There is a clear \hi\ detection in KUG~0203-Dw1, with morphology highly suggestive that the host galaxy KUG~0203-100 is actively stripping gas off of the dwarf, accompanied by a faint stellar stream. KDG~013 has a coincident \hi\ detection at its location, but the morphology of the gas indicates that it is unlikely to be associated with the UDG. While we consider that these UDGs may have formed from dwarf mergers, TDGs, or were preexisting UDGs subsequently processed by the group, the most likely formation origin is in typical dwarf galaxies that then suffered from tidal effects of their group, transforming them into UDGs. However, without additional follow-up such as kinematic measurements \citep[e.g.,][]{alabi2018}, it is impossible to completely rule out an accreted UDG origin. We favor the tidally ``puffed-up'' dwarf galaxy origin because the GC abundances in these UDGs are somewhat higher than what is found in gas bearing field UDGs, and this formation origin explains both the tidal features and the UDG nature of the objects without having to invoke multiple formation mechanisms.

These two UDGs complete a sample of five that were identified in the $\sim150^{2}$ deg CFHTLS-wide area. All of the UDGs were identified as having morphology suggestive of tidal interactions and in group environments of 2 or more. 
Our sample contains a dramatic outlier - UGC~9050-Dw1 - while the remaining four are somewhat self-consistent. The two UDGs out of the four (KUG~0203-Dw1 and KDG~013) that also have UV detections have slightly elevated GC populations compared to the two that do not have UV detections (NGC~2708-Dw1 and NGC~5631-Dw1), but all four have GC abundance expected for galaxies at their luminosity. Only one of these UDGs has a definitive \hi\ detection, which is actively being stripped from the galaxy (KUG~0203-Dw1). \citetalias{jones2021} concluded that NGC~2708-Dw1 and NGC~5631-Dw1 were also most consisted with tidal ``puffing'' within the group as their formation mechanism given the lack of neutral gas and the presence of the stellar streams, similar to our conclusions for KUG~0203-Dw1 and KDG~013. We speculate that if indeed all four of these UDGs formed from tidal effects, they may be at different evolutionary stages of the process. NGC~2708-Dw1 and NGC~5631-Dw1 were likely stripped the earliest (particularly NGC~5631-Dw1 with its anomalously low central surface brightness), given the lack of UV detections, lack of neutral gas, and the presence of longer lived stellar streams. This is followed by KDG~013, which was able to form a few more GCs, and lastly KUG~0203-Dw1 which has not been entirely stripped yet. 

However, one of these five UDGs is a significant outlier, displaying an exceptionally high GC abundance, a significant \hi\ mass, GALEX UV detection in the central region, and a significant tidal tail. \citetalias{fielder2023} contends that a dwarf major merger may be the most likely formation scenario for this particular UDG. Given this outlier and the general scatter of the observed properties of these five tidally influenced group UDGs, larger studies are warranted on both the modeling and observational front before we are able to draw broad conclusions about group UDGs displaying tidal features. 

Both N-body \citep{ogiya2018,benavides2021} and hydrodynamical simulations \citep{jiang2019,liao2019} have previously indicated that tidal stripping in groups is an important processing method for UDG formation. However, these simulations do not have predictions for GC abundances. Next generation hydrodynamical simulations will prove useful for making such comparisons. In addition, it is unclear how long stellar streams that some of these UDGs display may persist, since dwarf velocity dispersion, stellar radius and orbit eccentricity all play a role \citep[e.g.,][]{penarrubia2009}. Models of tidal features around UDGs interacting with massive galaxies will help us understand the fraction of the UDG population formed in this way.

There are several observational avenues that may prove fruitful for expanding the tidally influenced UDG sample, such as a manual inspection through the SMUDges catalog (Systematically Measuring Ultra-diffuse Galaxies; \citealt{zaritsky2019}) for objects with tidal features or even re-training the CNN (convolutional neural net) used for identifying objects in SMUDges to identify tidal features. Applying the source detection algorithm used to detect the UDGs studied here (see \citealt{bennet2017}) to other surveys (e.g., DECaLS; \citealt{dey2019}) may also yield additional sources. In the long term, such approaches could be extended to large-scale surveys like LSST. Future studies will allow for a more complete census of UDG formation in group environments (particularly those with tidal features) and allow us to more thoroughly assess the rarity of outlier UGC~9050-Dw1 and dwarf major mergers as a UDG formation mechanism in addition to understanding the fraction of UDGs formed by tidal processing in group environments.


\begin{acknowledgments}

This work is based on observations made with the NASA/ESA Hubble Space Telescope, obtained at the Space Telescope Science Institute, which is operated by the Association of Universities for Research in Astronomy, Inc., under NASA contract NAS5-26555.  These observations are associated with program \# 16890.

DJS acknowledges support from NSF grants AST-1821967, 1813708 and AST-2205863.

The work used data observed with the Karl G. Jansky Very Large Array. The National Radio Astronomy Observatory is a facility of the National Science Foundation operated under cooperative agreement by Associated Universities, Inc.

This work is based on observations obtained with MegaPrime/MegaCam, a joint project of CFHT and CEA/IRFU, at the Canada-France-Hawaii Telescope (CFHT) which is operated by the National Research Council (NRC) of Canada, the Institut National des Science de l'Univers of the Centre National de la Recherche Scientifique (CNRS) of France, and the University of Hawaii. This work is based in part on data products produced at Terapix available at the Canadian Astronomy Data Centre as part of the Canada-France-Hawaii Telescope Legacy Survey, a collaborative project of NRC and CNRS.

AK acknowledges financial support from the grant CEX2021-001131-S funded by MCIN/AEI/ 10.13039/501100011033 and from the grant POSTDOC\_21\_00845 funded by the Economic Transformation, Industry, Knowledge and Universities Council of the Regional Government of Andalusia. KS acknowledges support for the Natural Sciences and Engineering Research Council of Canada (NSERC).

This work was performed in part at the Aspen Center for Physics, which is supported by National Science Foundation grant PHY-2210452.

\end{acknowledgments}

%

\vspace{5mm}
\facilities{HST(ACS), VLA, CFHT}


\software{Astropy \citep{astropy2013,astropy2018,astropy2022}, 
          CASA \citep{mcmullin2007},
          CGAT-core \citep{CGAT-core},
          Dolphot \citep{dolphin2000,dolphin2016},
          Galfit \citep{peng2010},
          Photutils \citep{bradley2022},
          Reproject \citep{robitaille2020},
          Source Extractor \citep{bertin1996},    
          }



\appendix
\restartappendixnumbering
\section{Supplementary Tables}
\label{sec:appendix}

Here we include figures of the color-magnitude diagrams of sources identified by \textsc{DOLPHOT} in each UDG. We also provide data of the individual GC candidates identified in KUG~0203-Dw1 (\autoref{tab:kug0203Dw1_gccs}) and KDG~013 (\autoref{tab:kdg013_gccs}).

\begin{figure*}
    \centering
    \includegraphics[width=\columnwidth]
    {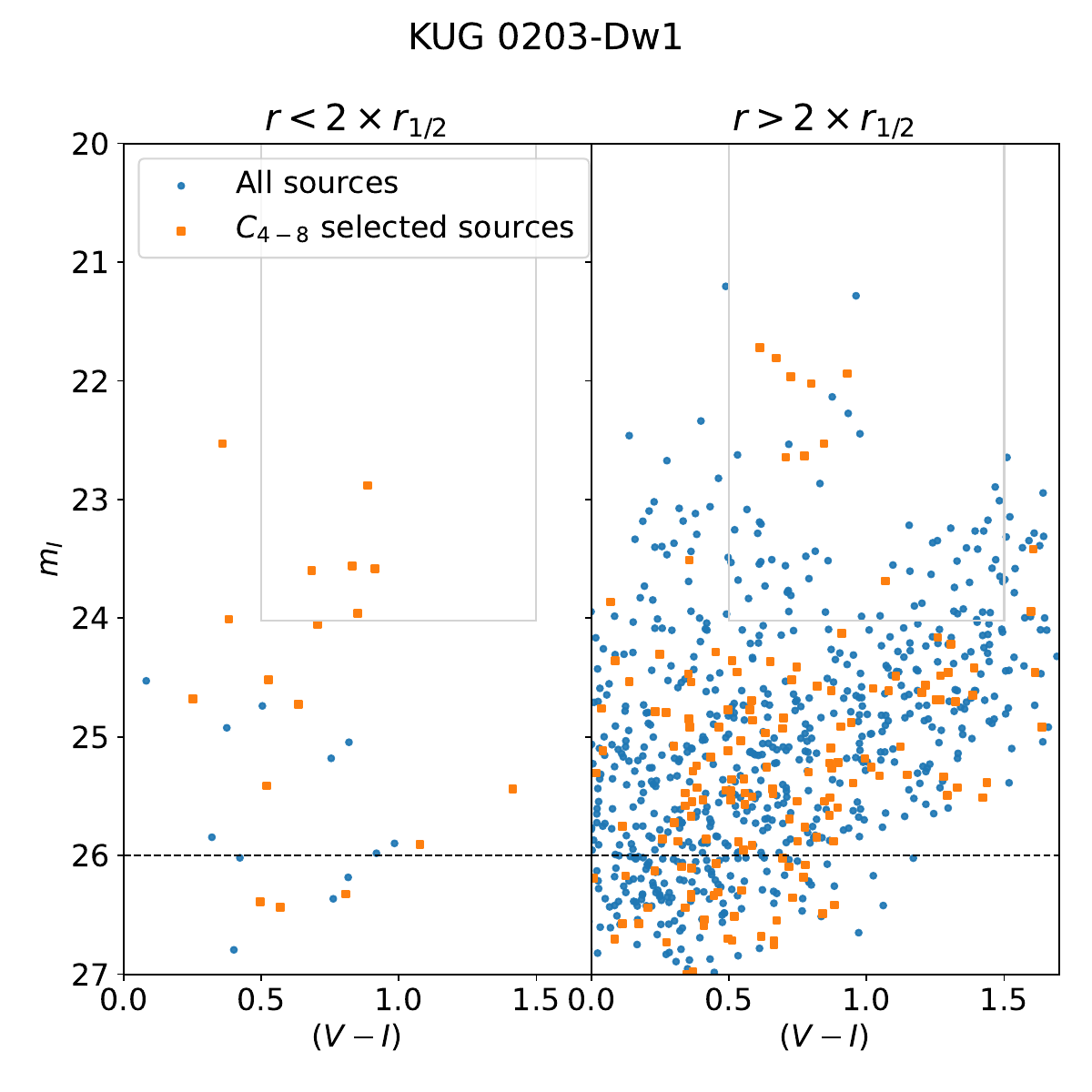}
    \includegraphics[width=\columnwidth]
    {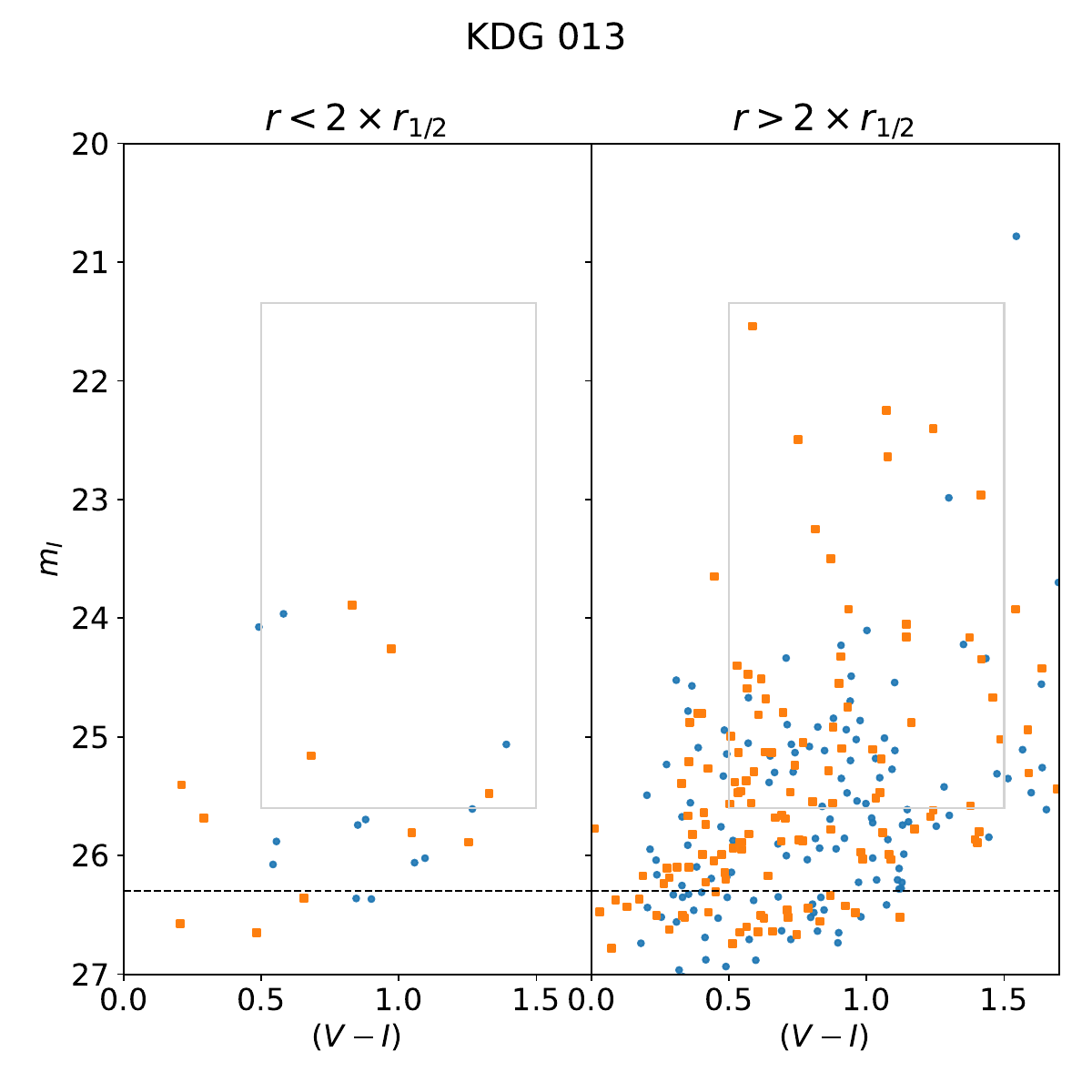}
    \caption{\textit{Left:} Color-Magnitude Diagrams (CMDs) of sources within the KUG~0203-Dw1 HST/ACS frame, presented in two panels: sources within the GC search radius of the UDG ($r<2\times r_{1/2}$) at left and sources outside of the search radius at right. Note that in the right panel we have excluded GC candidates that fall inside of the host galaxy KUG~0203-100 as they are unlikely contaminants. \textsc{Dolphot} identified sources classified as ``stars'' that meet the criteria for Signal-to-Noise (S/N), sharpness, roundness, and uncertainty described in \autoref{subsec:gc_selec} are represented by blue circles. Orange squares depict sources that successfully pass the subsequent concentration cut used to select GC candidates. The CMD space includes a gray outlined box representing the color and magnitude cuts for GCCs. A horizontal black dashed line indicates the 90\% completeness limit. \textit{Right:} Same as left, but for KDG~013.}
    \label{fig:CMD}
\end{figure*}

\begin{table*}
\centering
\caption{Globular cluster candidates in KUG~0203-Dw1}
    \hspace{-1.2in}
    \begin{tabular}{ccccccccccc}
    \hline\hline
    No. & R.A. & Dec &  F555W  &  F814W & $m_{V}$ & $m_{I}$ &  $M_{I}$ & $(V-I)$ & $c_{4-8}$ & $\rm{R}_{\rm{proj}}$ \\
    & deg & deg & mag & mag & mag & mag & mag & mag & mag & kpc \\
    \hline
    1 & 31.440109 & -9.836753 & $23.82 \pm 0.02$ & $22.88 \pm 0.01$ & 23.77 & 22.88 & -8.37 & 0.89 & 0.50 & 3.59 \\
    2 & 31.438186 & -9.843140 & $24.44 \pm 0.03$ & $23.56 \pm 0.02$ & 24.39 & 23.56 & -7.75 & 0.83 & 0.41 & 3.57 \\
    3 & 31.443381 & -9.836248 & $24.32 \pm 0.02$ & $23.60 \pm 0.02$ & 24.28 & 23.60 & -7.86 & 0.68 & 0.75 & 2.89 \\
    4 & 31.438003 & -9.846574 & $24.55 \pm 0.02$ & $23.58 \pm 0.01$ & 24.50 & 23.58 & -7.65 & 0.91 & 0.36 & 4.21 \\
    5 & 31.442585 & -9.842759 & $24.86 \pm 0.03$ & $23.96 \pm 0.02$ & 24.81 & 23.96 & -7.33 & 0.85 & 0.50 & 1.54 \\
    \hline
    \end{tabular}
    \\[4pt]
    Columns: 1) Number. 2) Right ascension in decimal degrees. 3) Declination in decimal degrees. 4) F555W apparent magnitude and errors determined from $\textsc{Dolphot}$. 5) F814W apparent magnitude and errors determined from $\textsc{Dolphot}$. 6) Extinction corrected $V$-band apparent magnitude using conversions presented in \citet{sirianni2005}. 7) Extinction corrected $I$-band apparent magnitude. 8) Extinction corrected $I$-band absolute magnitude. 9) $(V-I)$ color. 10) Concentration index determined from the GCC magnitudes in 4 and 8 pixel diameter apertures. 11) Projected distance from the center of KUG~0203-Dw1.
    \label{tab:kug0203Dw1_gccs}
\end{table*}

\begin{table*}[]
\centering
\caption{Globular cluster candidates in KDG~013}
    \hspace{-1.2in}
    \begin{tabular}{cccccccccccc}
    \hline\hline
    No. & R.A. & Dec &  F555W  &  F814W & $m_{V}$ & $m_{I}$ &  $M_{I}$ & $(V-I)$ & $c_{4-8}$ & $\rm{R}_{\rm{proj}}$ \\
     & deg & deg & mag & mag & mag & mag & mag & mag & mag & kpc \\
    \hline

    1 & 32.140328 & -7.847680 & $24.77 \pm 0.03$ & $23.89 \pm 0.02$ & 24.72 & 23.89 & -9.00 & 0.83 & 0.58 & 2.10 \\
    2 & 32.138603 & -7.853665 & $25.28 \pm 0.04$ & $24.26 \pm 0.02$ & 25.23 & 24.26 & -8.49 & 0.97 & 0.49 & 3.93 \\
    3 & 32.143510 & -7.845069 & $25.88 \pm 0.06$ & $25.16 \pm 0.04$ & 25.84 & 25.16 & -7.88 & 0.68 & 0.53 & 5.88 \\
    4 & 32.143052 & -7.844414 & $26.88 \pm 0.13$ & $25.48 \pm 0.06$ & 26.81 & 25.48 & -6.92 & 1.33 & 0.76 & 6.12 \\
    \hline
    \end{tabular}\\[4pt]
    Same as \autoref{tab:kug0203Dw1_gccs} but for KDG~013.
    \label{tab:kdg013_gccs}
\end{table*}

\bibliography{refs2,refs}{}
\bibliographystyle{aasjournal}



\end{document}